\documentclass[11pt]{article}

\usepackage{mathtools}
\usepackage{amssymb}
\usepackage{dsfont}
\usepackage{slashed}
\usepackage{fullpage}
\usepackage[colorlinks=true,linkcolor=blue,citecolor=magenta,linktocpage=true]{hyperref}
\usepackage{titlesec}
\titleformat*{\section}{\normalsize\bfseries}
\titleformat*{\subsection}{\normalsize\bfseries}
\titleformat*{\subsubsection}{\normalsize\bfseries}
\usepackage{cite}

\DeclareMathAlphabet{\bbvar}{U}{BOONDOX-ds}{m}{n}

\makeatletter
\renewcommand{\@dotsep}{10000}
\makeatother

\def\be#1\ee{\begin{align}#1\end{align}}
\def\nn{\nonumber}

\def\cC{\mathcal{C}}

\def\cE{\mathcal{E}}

\def\K{\mathcal{K}}

\def\cN{\mathcal{N}}

\def\cR{\mathcal{R}}

\def\V{\mathcal{V}}

\renewcommand{\sl}{{\mathfrak{sl}}}
\def\SL{\text{SL}}

\def\M{\mathcal{M}}

\renewcommand{\sl}{{\mathfrak{sl}}}

\def\rd{\textrm{d}}

\def\beq{\begin{eqnarray}}
\def\eeq{\end{eqnarray}}
\def\be{\begin{equation}}
\def\ee{\end{equation}}

\numberwithin{equation}{section}

\begin{document}

\title{\Large{\textbf{\sffamily Proper time reparametrization in cosmology: \\ M\"obius symmetry and Kodama charges}}}
\author{\sffamily Jibril Ben Achour\;$^{1, \;2} $}
\date{\small{\textit{$^{1}$ Arnold Sommerfeld Center for Theoretical Physics, Munich, Germany}  \\ \small{\textit{$^{2}$ Yukawa Center for Theoretical Physics, Kyoto University, Kyoto, Japan}}}}

\maketitle

\hrule
\hspace{.5cm}

\begin{abstract}
It has been noticed that for a large class of cosmological models, the gauge fixing of the time-reparametrization invariance does not completely fix the clock. Instead, the system enjoys a surprising residual Noether symmetry under a M\"{o}bius reparametrization of the proper time, which maps gauge-inequivalent solutions to the Friedman equations onto each other. In this work, we provide a unified treatment of this hidden conformal symmetry and its realization in the homogeneous and isotropic sector of the Einstein-Scalar-$\Lambda$ system. We consider the flat Friedmann-Robertson-Walker (FRW) model, the (A)dS cosmology and provide a first treatment of the model with spatial constant curvature.
We derive the general condition relating the choice of proper time and the conformal weight of the scale factor, and give a detailed analysis of the conserved Noether charges generating this physical symmetry. Our approach allows us to identify new realizations of this symmetry while recovering previous results in a unified manner.  We also present the general mapping onto the conformal particle and discuss the solution-generating nature of the transformations beyond the M\"{o}bius symmetry. Finally, we show that, at least in a restricted context, this hidden conformal symmetry is intimately related to the Kodama charges of spherically symmetric gravity. This new connection suggests that the M\"{o}bius invariance of cosmology is only the corner of a larger symmetry structure which could be relevant beyond cosmological models.
\end{abstract}
\hspace{.5cm}
\hrule

\thispagestyle{empty}
\newpage
\setcounter{page}{1}

\hrule
\tableofcontents
\vspace{0.7cm}
\hrule

\newpage

\section{Introduction}

 Symmetry reduced models of General Relativity (GR) play a pivotal role in our understanding of cosmological and black hole geometries. Investigating the gravitational dynamics in this simplified context, where a large number of degrees of freedom are frozen, turns out to have many advantages as one can have a better control on the resulting system to classify its solutions as well as to investigate its quantization. Moreover, the process of symmetry reduction can give rise to a rich underlying structure through the emergence of powerful enhanced symmetries \cite{Alekseev:2010mx}. In turn, such a hidden symmetry can give access to interesting solution-generating techniques to explore the solution space while revealing surprising properties of the underlying gravitational system.
 
 When restricting the gravitational field to a homogeneous geometry, the gauge invariance of the full theory reduces to a simple invariance under time-reparametrization. Fixing this gauge through a given choice of clock breaks the gauge symmetry and leaves us with a mechanical system for which the clock should, a priori, inherit the status of an absolute mechanical time. However, the situation turns out to be more subtle. For a large class of cosmological systems, the reduced action actually enjoys a residual Noether symmetry  under the 1d conformal group SL$(2,\mathbb{R})$ even after fixing the gauge. As a result, the clock is only defined as an equivalence class of clocks which can be mapped to each other under appropriate symmetry transformations. This residual symmetry is a global one, which acts as a M\"{o}bius reparametrization of the proper time. As a physical symmetry of the system, it maps gauge-inequivalent solutions to the Friedmann equations onto each other. The goal of this work is to provide a complete picture of this hidden symmetry for the class of homogenous and isotropic cosmological models, completing and generalizing previous findings.
 
 The existence of such residual conformal symmetry in cosmology was first noticed long time ago by Pioline and Waldron \cite{Pioline:2002qz}. It was observed that the dynamics of a (higher dimensional) vacuum (A)dS universe can be recast as the one of the conformal particle, an SO$(2,1)$ invariant mechanical system introduced in \cite{deAlfaro:1976vlx}. However, the mapping was restricted to a specific example and the explicit symmetries acting on the reduced cosmological action were not presented. Much later, an in-depth analysis of this conformal invariance was presented in \cite{BenAchour:2019ufa} in the context of the flat Friedmann-Robertson-Walker (FRW) model filled with a massless scalar field. See also \cite{BenAchour:2017qpb, BenAchour:2018jwq, BenAchour:2019ywl} for earlier investigations of this symmetry in the context of quantum cosmology. Starting from this analysis, the fate of this symmetry in cosmological systems was explored in different directions, revealing interesting generalizations and suggesting the existence of a larger underlying symmetry. On the one hand, the generalization of the $\SL(2,\mathbb{R})$ transformations to the whole Diff$(S^1)$ transformations was investigated in \cite{BenAchour:2020xif} and shown to provide a solution-generating map, using the property of the Schwarzian derivative, which allows one to reconstruct the (A)dS cosmology from the flat FLRW model. See also \cite{Gibbons:2014zla, Alon:2020yjq} for related discussions on this point. On the other hand, it was shown in \cite{BenAchour:2020njq, BenAchour:2020ewm} that the SL$(2,\mathbb{R})$ symmetry stands as a subsector of a larger SO$(3,2)$ group of symmetry which mixes the gravitational and matter degrees of freedom. Finally, this symmetry was found to hold for more general systems, such as the Schwarzschild-(A)dS black hole mechanics (described by the Kantowski-Sachs midi-superspace model) for which the symmetry is upgraded to the group SL$(2,\mathbb{R})\ltimes \mathbb{R}^3$\cite{Geiller:2020xze, Ben}. These different generalizations therefore suggest the existence of a deeper structure to be uncovered and beg for additional investigations. 
 
 In this work, we shall not focus on the recent developements mentioned above, but reconsider instead the class of homogeneous and isotropic cosmological systems. Indeed, despite the previous investigations, a complete picture of how this conformal symmetry is realized in such simple models is still missing and several questions have remained unanswered:
 \begin{itemize}
 \item Being realized as a M\"{o}bius reparametrization of the proper time, the symmetry discussed in \cite{BenAchour:2019ufa, BenAchour:2020xif} involves a gauge fixing. How does the choice of clock affect the realization of the residual symmetry ? What are the conditions selecting the clocks for which the symmetry is realized ?
 \item Moreover, under such a reparametrization of the proper time, the scale factor transforms as a primary field and not as a scalar. How does the conformal weight of the scale factor get modified when different clocks are used ? Can we derive some general conditions relating the choice of proper time and the value of the conformal weight ?
 \item Previous works have focused on the flat FRW cosmology as well as on its (A)dS extension \cite{BenAchour:2019ufa, BenAchour:2020xif}. In the framework of homogeneous and isotropic cosmological models, one can also consider models with a homogeneous spatial curvature being either positive, i.e. the closed universe, or negative, with hyperbolic spatial section. These models have not been treated yet. Do they also enjoy a residual symmetry similar to the flat and (A)dS models ? What is the fate of the symmetry when both sources of homogeneous curvature are turned on ? 
 \item Following the initial work by Pioline and Waldron \cite{Pioline:2002qz}, can we understand this symmetry through a unified mapping of these cosmological models into the conformal particle ?
 \item Finally, it is worth pointing out that up to now, the investigations on this residual conformal symmetry have been restricted to homogeneous gravitational fields describing either cosmology or the black hole interior. Can we find some hints towards a possible generalization to inhomogeneous spacetime such as inhomogeneous spherically symmetric gravity ?
 \end{itemize}   
 In the following, we shall address these questions step by step. We will consider the homogeneous and isotropic sector of the Einstein-Scalar-$\Lambda$ system and allow the geometry to admit a homogeneous constant spatial curvature. The model is therefore parametrized by two parameters $\left( L_c, L_{\Lambda}\right)$ which encode respectively the spatial constant curvature and the length scale associated to the cosmological constant. In order to investigate how the choice of proper time affects the realization of the symmetry, we futhermore introduce a one-parameter field dependent reparametrization of the lapse given by (\ref{para}) which encodes the freedom in choosing the clock. Starting from this general set-up, we compute the general variation of the cosmological action under a Diff$(S^1)$ transformation and identify the constraint relating the choice of proper time and the value of the conformal weight. Then, this general condition (\ref{cond}) allows us to identify in a unified manner the different realizations of the residual conformal symmetry. In addition to rederiving the known cases treated in \cite{BenAchour:2019ufa, BenAchour:2020xif}, we identify new realization of the symmetry for the flat FRW and (A)dS models, while providing a first treatment of the model with spatial constant curvature. This first part thus provides a unified treatment of this residual conformal symmetry in homogeneous and isotropic cosmological models. This is the first result of this work.
 
 In a second part,  we investigate the specific Diff$(S^1)$ transformation associated with a constant Schwarzian derivative known as Niederer's transformations \cite{Nied}. We show that such a conformal reparametrization is solution-generating: from the flat FRW model, they allow one not only to generate the potential term associated with the cosmological constant, but also the one associated with a constant spatial curvature. Thus, they provide non-trivial deformation of the CVH generators and allow one to turn on or off the potential contribution associated with a constant curvature in the symmetry reduced action, while preserving the $\sl(2,\mathbb{R})$ structure. This generalizes the findings presented in \cite{BenAchour:2020xif} for the (A)dS case. Then, we turn to the mapping onto the conformal particle. We show that the seminal work by Pioline is a subcase, associated to a specific choice of clock, belonging to a more general duality relating the different isotropic cosmological systems to the conformal particle with or without a Newton-Hooke interaction \cite{Galajinsky:2010ry}. We use this to comment on the possible consequences of such mapping at the quantum level.
 
In the third part of this work, we perform a detailed analysis of the conserved charges generating the M\"{o}bius invariance of each cosmological models. These charges are evolving constants of motion\footnote{The name was introduced in \cite{Evol}. In our case, it allows to stress the difference between the role of the CVH generators, which are not conserved along the evolution, and the Noether charges which are built as suitable time-dependent combinations of the CVH generators such that they are conserved. See \cite{Anderson:1995tu} for a discussion on the notion of evolving constant of motion.} built up from a set of phase space functions which form an $\sl(2,\mathbb{R})$ Lie algebra, known as the CVH algebra \cite{BenAchour:2017qpb, BenAchour:2018jwq, BenAchour:2019ywl}. We show explicitly how this charge algebra is modified when using different clocks to express the evolving constants of motion. Finally, we use them to solve the cosmological dynamics in an algebraic way, which provides an efficient shortcut to capture the cosmological dynamics, especially for the closed universe. 

The last part of this work aims at providing a first hint toward the last question raised above. Using the fact that our set-up is spherically symmetric, we compute the associated Kodama charges \cite{Kodama:1979vn}. The Kodama vector is a divergence-free vector field defined for any spherically symmetric background. Being divergence free, it naturally defines two conserved currents which are covariantly defined again for any spherically symmetric geometries. This fact is known as the Kodama's miracle and has triggered interesting investigations \cite{Racz:2005pm, Abreu:2010ru}. However, to our knowledge, the charges associated with these Kodama currents have not been analysed so far, and in particular the type of transformations they generate has remained elusive. We point, at least for a specific choice of clock and only for the flat FRW and (A)dS models, that the Kodama generators coincide with two of the phase space functions forming our CVH algebra, while the third piece is obtained through the Poisson bracket of the first two. It follows directly that the Kodama generators can be combined to form again a set of evolving constants of motion, which satisfies our time-dependent $\sl(2,\mathbb{R})$ Lie algebra. As a consequence, these conserved Kodama charges can be recognized as generating the residual symmetry of flat cosmology discussed in this work. To our knowledge, this is the first observation that the well known Kodama generators are associated with a conformal symmetry of the 
underlying gravitational system, at least in the specific example discussed here. From a more general perspective, this coincidence suggests that our symmetry might find some links with the volume-preserving diffeomorphism, to which the Kodama vector belongs. However, the result discussed here is restricted to a specific example and we have not been able to find a general mapping yet. More work would be required to go further. It nevertheless provides an interesting starting point to investigate possible connections beyond homogeneous spacetimes.

Having summarized our approach and our findings, let us briefly comment on the relevance of this symmetry for quantum cosmology. Independently of what should be the ultimate theory of quantum gravity, it is expected that cosmological geometries should emerge from some hydrodynamical approximation of this yet-to-be-constructed theory. Such emergent geometries would then result from a suitable mean field approximation of the quantum dynamics of the fundamental degrees of freedom of both quantum geometry and matter, capturing their collective behavior in the effective dynamics of a single collective wave function. This provides an interpretation for the wave function of the universe as obtained in the standard canonical Wheeler-de Witt approach. This point of view raised several key questions: given some candidate theory for quantum gravity, how one should select the suitable mean field approximation to extract semi-classical cosmology ? Can we find some criteria based on symmetry arguments ? We expect that the conformal symmetry discussed in this work might play a role in this direction. Concretely, our result might allow one to connect recent findings on quintic non-linear Schr\"{o}dinger equation and its conformal invariance, as discussed in \cite{Ghosh:2001an, Lidsey:2018byv}, with the current efforts in the group field theory approach attempting to extract cosmology as a quantum gravity condensate \cite{Gielen:2013naa, Oriti:2016qtz, Oriti:2016ueo, Oriti:2016acw, Pithis:2019tvp}. Another interesting road would be to follow the more developed fluid/gravity correspondence and seek for a dual description of cosmology in terms of conformal fluid dynamics \cite{Rangamani:2009xk}. Besides these on-going investigations, the residual symmetry discussed in this work might also be relevant for the problem of time in quantum cosmology \cite{Hoehn:2011jw}. Whether this residual symmetry under proper time reparametrization is only a corner of a much larger Noetherian symmetry describing internal change of clock, and associated with non-vanishing conserved charges, appears as a fascinating possibility. Current efforts are devoted towards exploring this possibility. Finally, we point out that the duality with the conformal particle opens the possibility to import and adapt results and methods developed for the quantization of the conformal particle to quantum cosmology, thus providing a new point of view to discuss long standing issues from a symmetry point of view.  See for examples along this line \cite{Ardon:2021vae, Khodaee:2017tbk, Andrzejewski:2015jya, Chamon:2011xk, Jackiw:2012ur, Camblong:2005an, Camblong:2003mz, Camblong:2003mb, Ananos:2003yh, Ananos:2002id}.

This work is organized as follows. Section~\ref{sec1} is devoted to describe our set-up. We present the Einstein-Scalar-$\Lambda$ reduced action and the geometry on which we shall focus. In Section~\ref{sec1.2}, we introduce our one-parameter field dependent reparametrization of the lapse and recast the action in the suitable form for our purpose. Section~\ref{sec2} is devoted to the unified treatment of the M\"{o}bius symmetry for all isotropic models and the derivation of the general conditions for this symmetry to be realized. In Section~\ref{sec2.4}, we present the solution-generating map based on the Niederer transformation, while Section~\ref{sec2.5} is devoted to the general mapping onto the conformal particle. Section~\ref{sec3} is devoted to a detailed analysis of the Noether charges, their algebra and the resolution of the dynamics. Finally, we discuss the relation to the Kodama generators in Section~\ref{sec4} and summarize our results and open directions in Section~\ref{sec5}.

\section{The cosmological model}

\label{sec1}

In this section, we specify the family of cosmological models we shall investigate in this work. They correspond to homogeneous and isotropic gravity filled with a massless scalar field and a cosmological constant. We present the symmetry reduced action and discuss the different choices of clocks to describe the dynamics. Finally, we briefly describe the phase space of our system before investigating its hidden symmetries. 

\subsection{Reduced action}
\label{sec1.1}

Consider the line element describing the homogeneous and isotropic curved FRW cosmological background given by
\be
\label{metric}
\rd s^2 = - N^2(t)\rd t^2 + a^2(t) \left[ \left( 1- \frac{r^2}{L^2_c} \right)^{-1}\rd r^2 + r^2 \rd \Omega^2\right] \;\;
\ee
where as usual, $N(t)$ is the lapse function, $a(t)$ is the scale factor and $L_c$ is the length scale encoding the constant spatial curvature of the background. One can switch from the closed universe with spherical spatial section to the open universe with hyperbolic spatial section by the mapping $L_{c} \rightarrow i L_c$. The flat case is recovered in the limit $L_c \rightarrow +\infty$. 
The symmetry reduced action of the Einstein-Scalar-$\Lambda$ system reads
\begin{align}
\label{ac}
S[g, \varphi]&  =  \int \rd^4 x \sqrt{|g|} \left[ \frac{\cR- 2 \Lambda}{L^2_P} - \frac{1}{2} g^{\mu\nu} \partial_{\mu}\phi \partial_{\nu} \phi \right] \\
&  = \frac{3\V_0}{L^2_P} \int  \rd t \left[ \frac{\left( a^2 a'' + a (a')^2 \right)}{N} - \frac{a^2 a' N'}{N^2}  +\frac{Na}{ L^2_c}  - \frac{ N a^3}{L^2_{\Lambda}}+ L^2_P \frac{a^3 (\phi')^2}{6N} \right] \;\;\;\; 
\end{align}
where a prime refers to a derivative w.r.t the time coordinate $t$. The length scale $L_P = \sqrt{8\pi G}$ is the Planck length, $L_{\Lambda} = \sqrt{3/2\Lambda}$ and we have introduced an homogeneous scalar field $\phi := \phi(t)$ without a self-interacting potential as a non-trivial matter source. Notice that we have restricted the spatial integration to a fiducial volume $\V_0$ given by
\be
\label{fidvol}
\V_0 = 4\pi \int^{r_{\ast}}_0 r^2 \left( 1- \frac{r^2}{L^2_c} \right)^{-1/2}  \rd r \;,
\ee
where $r_{\ast}$ sets the scale at which we restrict the integration in the radial direction. Thus $\V_0$ plays the role of the IR cut-off in our system and $[\V_0] = L^3$. We find convenient to introduce the rescaling
\be
\label{resc}
V_0 = 3 \V_0, \qquad \varphi = \phi /3, 
\ee 
Finally, integrating by part and upon implementing the above rescaling, the reduced action reads
\begin{align}
S[a, \varphi, N]& =  \frac{V_0}{L^2_P} \int  \rd t \left[ L^2_P \frac{a^3 (\varphi')^2}{2N}  - \frac{ a (a')^2}{N}  + \frac{Na}{ L^2_c}  -  \frac{N a^3}{L^2_{\Lambda}} +  \left( \frac{a^2 a'}{N}\right)' \right] 
\end{align}
Our goal is to investigate the realization of hidden symmetry of this cosmological system and in particular the role of proper time reparametrization. In that regard, it will reveal useful to find a suitable parametrization for the choice of proper time. This can be achieved by a field dependent reparametrization of the action which we now describe.

\subsection{Field dependent reparametrization}

\label{sec1.2}

 Consider the following field-dependent reparametrization where the lapse, and thus the time coordinate, are switched to
\be
\label{para}
N \rd t = \frac{\cN \rd \tau}{a^n}
\ee
where $n \in \mathbb{Z}$. The line element becomes
\be
\label{linerecast}
\rd s^2 = - \frac{\cN^2(\tau)}{a^{2n}(\tau)}\rd \tau^2 + a^2(\tau) \left[ \left( 1- \frac{r^2}{L^2_c} \right)^{-1}\rd r^2 + r^2 \rd \Omega^2\right]
\ee
and in term of this new time coordinate, the reduced action reads
\begin{align}
\label{met}
S_n [a, \varphi, \cN ]& =  \frac{V_0}{L^2_P} \int  \rd \tau \left[ L^2_P \frac{a^{3+n} \dot{\varphi}^2}{2\cN}  - \frac{ a^{1+n} \dot{a}^2}{\cN}  + \cN a^{1-n} \left(  \frac{1}{ L^2_c}   -  \frac{ a^{2}}{L^2_{\Lambda}} \right)+  \frac{\rd}{\rd \tau} \left( \frac{a^{3+n} \dot{a}}{\cN}\right)\right] 
\end{align}
where a dot refers to a derivative w.r.t. the new coordinate time $\tau$. Let us make two remarks at this stage. First, notice that the system is still gauge invariant and the gauge transformations are given by the time-reparametrization
\begin{align}
\label{gauge1}
\tau \rightarrow \tilde{\tau} & = f(\tau) \;, \\
\label{gauge2}
\cN \rightarrow \tilde{\cN}(\tilde{\tau}) & =  \dot{f}^{-1}(\tau)\cN(\tau) \\
\label{gauge3}
 a \rightarrow \tilde{a}(\tilde{\tau}) & = a(\tau) \\
 \label{gauge4}
 \varphi \rightarrow \tilde{\varphi}(\tilde{\tau}) & = \varphi(\tau)
\end{align}
which is the gauge invariance descending from the full diffeomorphism invariance of the theory after imposing homogeneity and isotropy. Finally, let us briefly discuss the notion of boundary in our set-up. Having consider a four-dimensional finite region of spacetime $\M$ and restricted the spatial integration to a fiducial cell with boundary $\partial \M$, we can split this boundary as $\partial \M = \Gamma \cup \Sigma_{+} \cup \Sigma_{-}$ where $\Gamma$ stands as the finite spatial boundary while $\Sigma_{\pm}$ are respectively the future/past spacelike boundaries. The assumption of homogeneity implies that all the fields evolve everywhere the same in that bulk region and on its boundary $\partial \M$. As a result, the only information about $\partial \M$ is contained in the fiducial volume $V_0$ which encodes the spatial integration. Therefore, in this homogeneous set-up, the boundary term reduces to a simple total time derivative for our mechanical action and it will not play any role in the following. However, notice that it should become important when relaxing homogeneity. 

Before diving in the exploration of the hidden symmetries of this system, let us write down its equations of motion. They reads
\begin{align}
\label{eom1}
\cE_{\cN} & = \frac{1}{\cN^2} \left[  a^{1 +n} \dot{a}^2 - \frac{L^2_P}{2} a^{3 +n} \dot{\varphi}^2 \right] + \frac{a^{1-n}}{L^2_c} - \frac{a^{3-n}}{L^2_{\Lambda}} \simeq 0 \\
\cE_{a} & = 2 \frac{\rd}{\rd \tau} \left( a^{1 +n} \frac{\dot{a}}{\cN}\right) - (\beta+1)^2 (1 +n) \frac{\dot{a}^2}{\cN} + (3+n) \; a^{2+n} \frac{L^2_P}{2} \frac{\dot{\varphi}^2}{\cN} \nn \\
\label{eom2}
& + \cN \left(\frac{1-n}{L^2_c} a^{-n} - \frac{3-n}{L^2_{\Lambda}} a^{2-n} \right) \simeq 0 \\
\label{eom4}
\cE_{\varphi} & = \frac{\rd}{\rd \tau} \left( a^{3+n} \frac{\dot{\varphi}}{\cN}\right) \simeq 0
\end{align}
The first equation (\ref{eom1}) is a constraint which enforces the invariance under time-reparametrization and forces the hamiltonian of the system to vanish. Together with the second equation (\ref{eom2}), they correspond to the Friedmann equations. The last equation (\ref{eom4}) is the continuity equation for the scalar matter field. Notice that depending on the choice of $n$, which encodes a choice of proper time to express the dynamics, the equations of motion can take very different forms as kinetic or potential terms can be removed. The question we shall investigate is whether this set of equations enjoy some hidden invariance for some specific choice of proper time, i.e. of parameter $n$. 
This question has been partially investigated in some specific cases in previous works, but a unified picture is still missing in the context of homogeneous and isotropic cosmological models. The goal of the next section is to derive the conditions of existence for these hidden symmetries for the family of models parametrized by $L_{c}$ and $L_{\Lambda}$. It will allow us to recover previous results found in \cite{BenAchour:2019ufa, BenAchour:2020xif} as well as to identify new realizations of these hidden symmetries, in particular in the case of the closed universe which is treated for the first time here.
 
\subsection{Hamiltonian formulation}

\label{sec1.3}

Before discussing the hidden symmetries, let us complete the presentation of our cosmological system by writing down its phase space. The conjugated momenta to the gravitational and matter configuration d.o.f $(a, \varphi)$ are given by
\be
\pi_a = - \frac{2V_0}{L^2_P} \frac{a^{1+n} \dot{a}}{\cN} \;, \qquad \pi_{\varphi} = V_0 a^{3+n} \frac{\dot{\varphi}}{\cN}
\ee
such that $[\pi_a] = [a] = 1$, while $[\pi_{\varphi} ] =L$ and  $[\varphi] = L^{-1}$ and we have
\be
\{ a, \pi_a\} = \{ \varphi, \pi_{\varphi} \} = 1
\ee
The hamiltonian reads
\begin{align}
H_n[\cN] & 
=   \frac{\cN}{V_0} \left[ \frac{\pi^2_{\varphi}}{2a ^{3+n}} - \frac{L^2_P}{4} \frac{\pi^2_a}{a^{1+n}} - \frac{ V^2_0 }{ L^2_P L^2_c} a^{1-n} + \frac{V^2_0}{ L^2_PL^2_{\Lambda}}a^{3-n} \right] \simeq 0
\end{align}
Notice that we have $[H] = L^{-1}$. In the following, it will be useful to introduce the alternative canonical pair
\be
v = a^{3+n} \;, \qquad \pi_v = \frac{1}{(3+n)}\frac{\pi_a}{a^{2+n}} \;, \qquad \text{such that} \qquad \{ v, \pi_v\} = 1
\ee
where we have assumed $n\neq -3$. This choice of clock corresponds to setting the volume to unity and therefore to select the unimodular clock. In term of the new canonical variables, the hamiltonian reads
\begin{align}
H_n[\cN] & 
=   \frac{\cN}{V_0} \left[ \frac{\pi^2_{\varphi}}{2v} - \frac{(3+n)^2}{4} L^2_P v \pi^2_v  -  \frac{ V^2_0 }{ L^2_P L^2_c} v^{\beta_1} + \frac{V^2_0}{ L^2_PL^2_{\Lambda}} v^{\beta_2} \right] \simeq 0
\end{align}
where 
\be
\beta_1 = \frac{1-n}{3+n} \;, \qquad \beta_2 = \frac{3-n}{3+n}
\ee
It will also reveal useful to define the dilatation generator as
\be
\label{dil}
\cC_n : =   \frac{2 V_0 }{(3+n)^2 L^2_P} \; \{ v, H_n \} =  -  v \pi_v
\ee
which coincides with the (rescaled) speed of $v$.

\section{Diff$(S^1)$ proper time reparametrization of the action}

\label{sec2}

The conformal symmetry found in \cite{BenAchour:2019ufa, BenAchour:2020xif} is realized as suitable reparametrization of the proper time under which the gravitational configuration variable transforms as a primary field. Following these works, we introduce the proper time coordinate
\be
\rd \eta = \cN \rd \tau
\ee
and we write the reduced action as
\begin{align}
\label{acc}
S_n [a, \varphi]& =  \frac{V_0}{L^2_P} \int  \rd \eta \left[ \frac{L^2_P}{2} a^{3+n} \dot{\varphi}^2  -  a^{1+n} \dot{a}^2  +  a^{1-n} \left(  \frac{1}{ L^2_c}   -  \frac{ a^{2}}{L^2_{\Lambda}} \right)\right] 
\end{align}
where we use again a dot to refer now to the time derivative w.r.t the proper time coordinate $\eta$.

Writing the action in term of the proper time $\eta$ has important consequences on the underlying symmetry of the system. The gauge invariance under time-reparametrization (\ref{gauge1}-\ref{gauge4}) which requires the presence of the lapse is no longer realized such that the equation of motion (\ref{eom1}) is not imposed anymore as a first class constraint. As it stands, the action (\ref{acc}) encodes the dynamics of a mechanical system whose hamiltonian can have any positive real values, i.e. $H = E $. In order to encode the same dynamics as before, we have to restrict the system such that its hamiltonian vanishes. The only difference is that now, this constraint is not enforced dynamically but fixed by hand. Keeping in mind this point, the action (\ref{ac}) supplemented with the restriction $H=0$ nevertheless encodes the very same cosmological dynamics as before. At this stage, one might misleadingly conclude that the gauge invariance (\ref{gauge1}-\ref{gauge4}) being broken, the proper time $\eta$ stands as an absolute time. The situation is however more subtle as the action (\ref{ac}) turns out to enjoy some surprising hidden symmetries. 

In order to see this, consider a general Diff$(S^1)$ transformation given by
\begin{align}
\label{trans1}
\eta \rightarrow \tilde{\eta} &= f(\eta) \\
\label{trans2}
  \varphi \rightarrow \tilde{\varphi}(\tilde{\eta}) &=  \varphi(\eta) \\
  \label{trans3}
   a \rightarrow \tilde{a}(\tilde{\eta}) &= \left[ \dot{f}(\eta) \right]^{\lambda}a(\eta) 
\end{align} 
such that the scalar field $\varphi$ transforms as a scalar while the scale factor $a$ transforms as a primary field with conformal weight $\lambda$. Notice that the transformation of the scale factor (\ref{trans3}) is rather different from its standard behavior under standard time-reparametrization given by (\ref{gauge3}) where it transforms instead as a scalar. Performing this finite Diff$(S^1)$ transformation on the cosmological action, one can show that provided
\be
\label{cond}
\lambda \left( 3+n \right)  -1 = 0 \qquad \text{and} \qquad n \neq - 3
\ee
the variation of the cosmological action reads
\begin{align}
\label{main}
\Delta S_n & = -  \frac{V_0}{L^2_P} \int \rd \eta \left[2 \lambda^2 \frac{\rd}{\rd \eta} \left( \frac{\ddot{f}}{\dot{f}} a^{3+n}\right) \right. \\
\label{change}
& \left. \qquad \qquad \qquad - 2 \lambda^2 \text{Sch}[f]  a^{3+n} - \frac{a^{1-n}}{L^2_c} \left( \dot{f}^{2-2\lambda ( n+1)} -1 \right) +  \frac{a^{3-n}}{L^2_{\Lambda}} \left( \dot{f}^{2- 2\lambda n} -1 \right) \right]
\end{align}
where we have introduced the Schwarzian derivative of the function $f(\eta)$ defined as
\be
\text{Sch}[f]:= \frac{\dddot{f}}{\dot{f}} - \frac{3}{2} \left( \frac{\ddot{f}}{\dot{f}}\right)^2 
\ee 
This variation is the first result of this work. It allows to identify in a clean way the hidden symmetries of the cosmological action. The first term is a total derivative term which does not modify the dynamics. The second term encodes the conformal anomaly of the transformation, given by the Schwarzian cocycle. As it does not depend on the length scale $(L_c, L_{\Lambda})$, this term will survive in the flat case. We shall refer to it as the flat cosmological anomaly. The two last terms encode respectively the corrections to the flat anomaly coming from i) the constant curvature of the spatial section given by $1/L^2_c$ or ii) the presence of a non-vanishing cosmological constant $1/L^2_{\Lambda}$. Notice that the condition (\ref{cond}) shows that the choice of proper time $\eta$ dictates the value of the conformal weight $\lambda$ entering in the transformation (\ref{trans3}). Using the general result (\ref{main}), we shall now identify the possible hidden symmetries associated to the different families of cosmological models parametrized by the length scales $(L_c, L_{\Lambda})$. 

\subsection{Symmetries}

As a first result, notice that when both $(L_c, L_{\Lambda})$ are present, no symmetry can be realized. This is manifest from the fact that there is no choice for $n$ which would allow one to factorized the powers of the scale factor involved in the last three terms of (\ref{main}). Let us now discuss the different families of cosmological models.

\subsubsection{Flat FRW cosmology}

\label{sec2.1}

The flat FRW cosmology is obtained in the limit $L_c \rightarrow + \infty$ and $L_{\Lambda} \rightarrow + \infty$. In that case, we see that the variation of the action under a Diff$(S^1)$ transformation reduces to
\begin{align}
\label{mainflat}
\Delta S_n & = \frac{2 \lambda^2 V_0}{L^2_P} \int \rd \eta  \left[   \text{Sch}[f]  a^{3+n}  - \frac{\rd}{\rd \eta} \left( \frac{\ddot{f}}{\dot{f}} a^{3+n}\right) \right]
\end{align}
This result shows that at each value of $n \in \mathbb{Z}/\{-3\}$, there is a solution for $\lambda$, given by (\ref{cond})
such that the transformations satisfying $ \text{Sch}[f] =0$ are Noether symmetries of our cosmological system, i.e the action varies only by a total derivative term. These transformations are given by
\be
\label{mobius}
f(\eta) = \frac{a \eta + b}{c\eta + d} \;, \qquad \text{with} \qquad ad - bc \neq 0
\ee
where $(a,b,c,d)$ are real constant parameters. One can show that among these transformations, the non trivial ones can be accounted by considering the subset satisfying $ad-bc =1$ while quotienting by the transformations $(a,d,b,c) \rightarrow (-a,-b,-c,-d)$. As a result, the symmetry group is given by SL$(2,\mathbb{R})/\mathbb{Z}_2$ which corresponds to the group of M\"{o}bius transformations.
This provides an extension of the result found in \cite{BenAchour:2019ufa} which corresponds to $n=0$ and $\lambda = 1/3$. Before going to the next model, we point that the specific value $n=-3$ is forbidden because the transformation generates a global factor which cannot be removed by any choice of the conformal weight, such that this case does not enjoy any symmetry. Let us now investigate the case where a source of constant curvature is turned on.

\subsubsection{(A)dS cosmology}

\label{sec2.2}

Consider the (A)dS cosmological system which corresponds to $L_c\rightarrow + \infty$ and let us focus on the de Sitter model with $L_{\Lambda} >0$. The AdS case can be obtained by sending $L_{\Lambda} \rightarrow i L_{\Lambda}$. The transformation of the reduced action becomes
\begin{align}
\label{main}
\Delta S_n & = \frac{V_0}{L^2_P} \int \rd \eta \left[ 2 \lambda^2  \text{Sch}[f] a^{3+n}  -  \frac{a^{3-n}}{L^2_{\Lambda}} \left( \dot{f}^{2 - 2 \lambda n} -1 \right) - 2 \lambda^2 \frac{\rd}{\rd \eta} \left( \frac{\ddot{f}}{\dot{f}} a^{3+n}\right)\right]
\end{align}
In general, the second term prevents the symmetry to be realized. However, there are two special values for $n$ for which a symmetry emerges.  
\\ \\ 
\textbf{Case 1:} A first possibility is realized for the special case $n=3$ which imposes $\lambda = 1/6$. In that case, the contribution of the cosmological constant reduces to a pure constant in the initial lagrangian, similar to the choice used for the Kantowschi-Sachs model in \cite{Geiller:2020xze}. Then, the variation of the action reads
\begin{align}
\Delta S_3 & = \frac{V_0}{3L^2_P} \int \rd \eta \left[  \text{Sch}[f] a^6 - \frac{\rd}{\rd \eta} \left( \frac{\ddot{f}}{\dot{f}} a^6 - \frac{3(f- \eta)}{L^2_{\Lambda}}\right)\right]
\end{align}
The Noether symmetry is then realized by imposing only $ \{ f, \eta \} =0$ which coincides with the condition imposed in the flat case, i.e.
\be
\label{mobius}
f(\eta) = \frac{a \eta + b}{c\eta + d} \;, \qquad \text{with} \qquad ad - bc \neq 0
\ee
This alternative realization of the symmetry provides a new result for (A)dS cosmology and shows how the M\"{o}bius invariance of the system can be realized in different forms depending on the choice of the clock.
\\ \\
\textbf{Case 2:} For $n=0$, one obtains from (\ref{cond}) that $\lambda = 1/3$. The volume $a^3$ entering in the two first terms can then be factorized and the variation of the action becomes
\begin{align}
\Delta S_0 & = \frac{2V_0}{9L^2_P} \int \rd \eta \left[ \left(  \text{Sch}[f]  -  \frac{9}{2L^2_{\Lambda}} \left( \dot{f}^{2} -1 \right) \right) a^3 -  \frac{\rd}{\rd \eta} \left( \frac{\ddot{f}}{\dot{f}} a^{3}\right)\right]
\end{align}
This variation reduces to a pure total derivative term and thus to a Noether symmetry of the system provided $f(\eta)$ satisfies the differential equation
\be
\text{Sch}[f]  -  \Omega_{\Lambda}^2 \dot{f}^{2}  = -  \Omega_{\Lambda}^2 \;, \qquad \Omega_{\Lambda} = \frac{3}{\sqrt{2}L_{\Lambda}}
\ee
This differential equation can be recast using the composition law for the Schwarzian. See Appendix~\ref{app} for detail. The reparametrization function $f(\eta)$ solving this differential equation is given by
\be
\label{defmob}
f(\eta) = F^{-1} \circ M \circ F \;, \qquad \text{with} \qquad M := \frac{a \eta + b}{c \eta +d} \;, \qquad F:= \tanh{\left[ \frac{\Omega_{\Lambda} \eta}{\sqrt{2}}\right]}
\ee
The conserved charges and the transformations of the solution space of this system have been discussed previously and we shall not expand further on this specific case in what follows. See \cite{BenAchour:2020xif} for details.

\subsubsection{The closed universe}

\label{sec2.3}

Finally, we consider the closed universe which has not been treated yet.  The closed universe is obtained by taking the limit $L_{\Lambda} \rightarrow +\infty$ and choosing $L_c >0$ (The model with homogeneously curved hyperbolic spatial section is obtained by the mapping $L_c \rightarrow i L_c$). For this system, the transformation of the action reduces to
\begin{align}
\Delta S_n & = \frac{V_0}{L^2_P} \int \rd \eta \left[ 2 \lambda^2 \text{Sch}[f] a^{3+n} + \frac{a^{1-n}}{L^2_c} \left( \dot{f}^{2-2\lambda ( n+1) } -1 \right)  - 2 \lambda^2 \frac{\rd}{\rd \eta} \left( \frac{\ddot{f}}{\dot{f}} a^{3+n}\right)\right]
\end{align}
There are again two special values for $n$ for which a Noether symmetry is realized:
\\ \\ 
\textbf{Case 1:} A first possibility corresponds to $n=1$ which implies $\lambda = 1/4$. This choice allows one to recast the contribution from the constant curvature to a pure constant in the action. The transformation reduces then to
\begin{align}
\Delta S_1 & = \frac{V_0}{8L^2_P} \int \rd \eta \left[ \text{Sch}[f] a^{4}  -  \frac{\rd}{\rd \eta} \left( \frac{\ddot{f}}{\dot{f}} a^{4}  +   \frac{8( f -\eta)}{L^2_c}  \right)\right]
\end{align}
and the symmetry is again obtained by imposing $\text{Sch}[f] =0$, selecting the M\"{o}bius transformation (\ref{mobius}). This choice parallels the second possibility found for the (A)dS case.
\\ \\
\textbf{Case 2:} The second possibility is to choose $n= -1$ which imposes $\lambda = 1/2$. Then, the variation of the action becomes
\begin{align}
\Delta S_{-1} & = \frac{ V_0}{2 L^2_P} \int \rd \eta \left[ \left(  \text{Sch}[f] +  \frac{2}{L^2_c} ( \dot{f}^{2} -1) \right) a^2  -  \frac{\rd}{\rd \eta} \left( \frac{\ddot{f}}{\dot{f}} a^{2}\right)\right]
\end{align}
Interestingly, the condition to have a Noether symmetry is the same as for the (A)dS case up to a sign and the definition of the pulsation. It reads
\be
\label{eqdiff}
\text{Sch}[f]+  \Omega^2_c \dot{f}^{2}  = \Omega^2_c \;, \qquad \Omega_c = \frac{\sqrt{2}}{L_{c}}
\ee
The resolution of the third order differential equation is discussed in Appendix~\ref{app}.
Therefore the closed universe enjoys a M\"{o}bius invariance under the reparametrization
\be
\label{comp}
f(\eta) = F^{-1} \circ M \circ F \;, \qquad \text{with} \qquad M := \frac{a \eta + b}{c \eta +d} \;, \qquad F:= \tan{\left[ \frac{\Omega_c \eta}{\sqrt{2}}\right]}
\ee 
It is intriguing to see that the closed universe and the (A)dS cosmology, despite encoding radically different dynamics, share the very same symmetry, provided we identify the right clocks. This is made possible by the specific values of the parameters $(n, \lambda)$ allowing one to recast the reduced action for the closed universe model in a form similar to the (A)dS case.
\\ \\
Let us now summarize our findings. We have shown that under a general Diff$(S^1)$ transformation in proper time, the cosmological action parametrized by the two length scales $(L_c, L_{\Lambda})$ transforms as (\ref{main}). This result has allowed us to identify in a unified manner the hidden symmetries of the different cosmological models of homogeneus and isotropic Einstein-Scalar-$\Lambda$ gravity: the flat FRW cosmology, the (A)dS cosmology and finally to treat the closed universe model:
\begin{itemize}
\item We have recovered known results presented in \cite{BenAchour:2019ufa} and \cite{BenAchour:2020xif}, namely the M\"{o}bius invariance of the flat model for $\left( n, \lambda \right) = (0, 1/3)$ and the one of (A)dS cosmology for $\left( n, \lambda \right) = (0, 1/3)$.  
\item We have also obtained new results for these two models : namely that the M\"{o}bius invariance of the flat model is realized for any choice of proper time, i.e. for any choice of $n \in \mathbb{Z}/\{-3 \}$. Moreover, we have identified a second realization of the M\"{o}bius symmetry for (A)dS cosmology given by $\left( n, \lambda \right) = (3, 1/6)$. 
\item  Finally, we have treated for the first time the spatially homogeneously curved model. We have shown that this model admits also two specific choices of proper time for which a M\"{o}bius invariance is realized, namely $(n, \lambda) = (1, 1/2)$ and $(n, \lambda) = (-1, 1/4)$.
\end{itemize}
Therefore, we have shown that, upon identifying the suitable proper time, the above homogeneous and isotropic cosmological systems enjoy a hidden symmetry under the one dimensional conformal group $\SL(2,\mathbb{R}) $. This symmetry is residual in the sense that it exists after having gauge fixed the time-reparametrization invariance of the system. The existence of this global Noether symmetry provides thus a second layer of relativeness in the status of the clock used to follow the cosmological dynamics which can thus be defined only as an equivalence class of 
clocks.

Let us now make few remarks regarding the more general Diff$(S^1)$ transformations. While they are not symmetries of the action, they nevertheless play an interesting role. Under these transformations, the action gets shifted by the standard cocycle term given by the Schwarzian of the reparametrization function. This cocyle governs the conformal anomaly of the action under these general conformal transformations and the constant term in front of it can be interpreted as a central charge. In order to be dimensionless, one can rescale the time coordinate such that $\eta \rightarrow \eta /L_{P}$. One can then associate to the cosmological system a central charge given by
\be
c = \frac{V_0}{L^3_P}
\ee
which encodes the ratio between the IR cut-off of our system, namely the fiducial volume $V_0$, and the UV cut-off given by the Planck length $L_P$. This result fits with the standard interpretation of the central charge as encoding the effective number of degrees of freedom of the system. One can therefore view the Schwarzian as a new interaction term and the central charge as its coupling constant. While in general, this new contribution to the potential is time-dependent, one can select a subset of Diff$(S^1)$ transformations which generate a constant potential such that it allows to generate the contribution of a constant curvature in the reduced action. Let us demonstrate this explicitly.

\subsection{Conformal bridge, Niederer transformation and Schwarzian curvature}

\label{sec2.4}

\label{Niederer}
In this section, we show that, starting from the flat model, one can generate the potential terms encoding the presence of a homogeneous curvature using suitable Diff$(S^1)$ transformation. This mapping is known as the Niederer transformations \cite{Nied}. Consider thus the reduced action for the flat model written in proper time
\begin{align}
S_n [a, \varphi]& =  \frac{V_0}{L^2_P} \int  \rd \eta \left[ \frac{L^2_P}{2} a^{3+n} \dot{\varphi}^2  -  a^{1+n} \dot{a}^2 \right] 
\end{align}
The Niederer's transformation are given by
\begin{align}
\label{NT0}
\eta \rightarrow \tilde{\eta} & = \frac{\sqrt{2}}{\Omega} \tan{\left( \frac{\Omega \eta}{\sqrt{2}}\right)} \\
\label{NT1}
a \rightarrow \tilde{a} & = \frac{a}{\left[ \cos{\left( \frac{\Omega \eta}{\sqrt{2}}\right)}\right]^{2\lambda}} \\
\label{NT2}
\varphi \rightarrow \tilde{\varphi} & = \varphi 
\end{align}
It is easy to see that if $\Omega \in \mathbb{R}^{\ast}$, the transformation (\ref{NT0}) compactifies the clock $\tilde{\eta} \in ] -\infty, +\infty [$ to $\eta \in ]-\pi/(\sqrt{2}\Omega), \pi/(\sqrt{2}\Omega) [$.  However, if $\Omega  \in \mathbb{R}^{\ast} $, this map decompactifies instead the clock $\tilde{\eta}$. As we shall see, this feature reflects a physical property of the cosmological models generated by this transformation. 

Now, notice that the above transformation belongs to the Diff$(S^1)$ transformation with $f(\eta) =\sqrt{2} \tan{[\Omega \eta/\sqrt{2}]}/ \Omega$. Using that the Schwarzian derivative of this function is a constant, i.e.
\be
\{ f, \eta\} = \Omega^2
\ee
the general variation of the action (\ref{mainflat}) gives us
\begin{align}
\tilde{S}_n =  \frac{V_0}{L^2_P} \int  \rd \eta \left[ \frac{L^2_P}{2} a^{3+n} \dot{\varphi}^2  -  a^{1+n} \dot{a}^2 + \frac{2 \Omega^2}{(3+n)^2} a^{3+n}  - \frac{\rd}{\rd \eta} \left( \frac{\ddot{f}}{\dot{f}} a^{3+n}\right)\right]
\end{align}
The Niederer's transformation generates therefore a potential term linear in $a^{3+n}$ which modifies the dynamics. This transformation allows one to reconstruct the (A)dS cosmological system as well as the model with homogeneous spatial curvature. To see that, consider the two following cases:
\begin{itemize}
\item For $(n, \lambda) = (0, 1/3)$, and upon setting 
\be
\Omega =  \left\{
    \begin{array}{ll}
        i \Omega_{\Lambda}  & \mbox{for de Sitter cosmology $\Lambda >0$} \;   \\
        \Omega_{\Lambda}   & \mbox{for Anti de Sitter cosmology $\Lambda >0$}  \\
    \end{array}
\right.
\ee
where $\Omega_{\Lambda} = 3/ (\sqrt{2}L_{\Lambda})$, the resulting action becomes
\begin{align}
\tilde{S}_0 =  \frac{V_0}{L^2_P} \int  \rd \eta \left[ \frac{L^2_P}{2} a^{3} \dot{\varphi}^2  -  a \dot{a}^2 -  \frac{a^{3}}{L^2_{\Lambda}} \right]
\end{align}
which corresponds to the (A)dS cosmological action in proper time.  
\item For $(n, \lambda) = (-1, 1/2)$, and upon setting 
\be
\Omega =  \left\{
    \begin{array}{ll}
         \Omega_{c}  & \mbox{for closed universe}   \\
        i\Omega_{c}   & \mbox{for hyperbolic universe}  \\
    \end{array}
\right.
\ee
where $\Omega_{\Lambda} = \sqrt{2}/ L_c$, the resulting action becomes
\begin{align}
\tilde{S}_0 =  \frac{V_0}{L^2_P} \int  \rd \eta \left[ \frac{L^2_P}{2} a^{3} \dot{\varphi}^2  -  a \dot{a}^2 +  \frac{a}{L^2_c} \right]
\end{align}
which corresponds to the reduced action for the closed universe in proper time.  
\end{itemize}
Therefore, the Niederer's transformation transforms the reduced action of the flat cosmological model to either the (A)dS action or to the universe with homogeneous spatial curvature. It follows that the equations of motion for these systems are also mapped to each other, providing therefore a solution-generating method for these family of cosmological models. Moreover, the compactifiction of the time coordinate $\tilde{\eta}$ of the flat model when $\Omega \in \mathbb{R}^{\ast}$ can be understood from the fact that the curvature can stop the cosmic expansion in AdS cosmology and in the closed universe model, such that the universe in both models reaches a finite size which can be monitored by a compact time coordinate $\eta$. 

While this result was already known for the (A)dS case \cite{BenAchour:2020xif}, the present treatment shows that it also generalizes to the case where a spatial constant curvature is turned on. Interestingly, the two sources of homogeneous curvatures, despite having radically different effects on the cosmological dynamics, can be both encoded using a constant Schwarzian cocyle. Moreover, in general, this Schwarzian term is a time-dependent function and for $n=0$, the new interaction term comes with the right power of the scale factor, i.e. $a^3$, to mimic a self-interaction potential for the scalar field. Whether this Schwarzian can be used to generate inflationary dynamics out of the free case is an interesting direction which begs for more investigations. See \cite{Lidsey:2018byv} for interesting results in this direction.

Coming back to the case of a constant Schwarzian transformation discussed here, it is interesting to note that such mappings are in close analogy with the one discussed for example in \cite{Andrzejewski:2014boa, Galajinsky:2015xla} which relates the conformal particle with its Newton-Hooke version. This analogy is actually not a mere accident, as one can show that the isotropic cosmological models discussed here can be recast into the conformal particle using different appropriate clocks. This is the subject of the next section.

\subsection{Mapping on the conformal particle}

\label{sec2.5}

Now, let us show how the different cosmological models can be recast, upon using the right choice of clock, as a conformal particle with or without a Newton-Hooke interaction \cite{deAlfaro:1976vlx}. The first step is to recast the kinetic matter term as the potential contribution of the action. The reason is that for a scalar field without self-interacting potential, one has $\dot{\pi}_{\varphi} =0$. Performing a Legendre transform on the matter sector, we obtain
\begin{align}
\label{LT}
\bar{S} [a]& =\int \rd \eta \pi_{\varphi} \dot{\varphi} - S_n [a, \varphi ] \\
& =  \frac{V_0}{L^2_P} \int  \rd \eta \left[ a^{1+n} \dot{a}^2  + \frac{L^2_P}{2V^2_0} \frac{\pi^2_{\varphi}}{a^{3+n}} + \frac{a^{3-n}}{L^2_{\Lambda}} -  \frac{a^{1-n}}{ L^2_c}  \right] 
\end{align}
where we have ignored the boundary term.  At this stage, one can derive different mapping onto the conformal particle depending on the choice of clock, namely on the value of the parameter $n$. First, we introduce the following configuration variable
\be
q (\eta)= \frac{2}{(3+n)} \sqrt{a^{3+n}(\eta)} \;, \qquad \frac{\rd q}{\rd \eta} = \sqrt{a^{1+n}} \frac{\rd a}{\rd \eta}
\ee
together with a coupling constant, which also plays the role of the $\sl(2,\mathbb{R})$ Casimir of the conformal particle, given by
\be
\mathcal{G} = -  \frac{2L^2_P \pi^2_{\varphi}}{(3+n)^2V^2_0}
\ee
Using these redefinitions, a first possibility consists in recasting the cosmological action as 
\begin{align}
\label{accqm1}
\bar{S} [q]& =  \frac{V_0}{L^2_P} \int  \rd \eta \left[ \dot{q}^2 - \frac{\mathcal{G}}{q^2} + \Omega^2 q^2 \right] 
\end{align}
where we have
\begin{align}
\Omega = \left\{
    \begin{array}{ll}
        \frac{3}{2L_{\Lambda}} & \mbox{if } L_c \rightarrow + \infty \;, \;\;n=0 \\
        \frac{1}{L_{c}} & \mbox{if } L_{\Lambda} \rightarrow + \infty \;, \;\;n= -1 \\
        0 & \mbox{if } L_{\Lambda} , L_{c} \rightarrow + \infty \;, n \in \mathbb{Z} \\
    \end{array}
\right.
\end{align}
The action (\ref{accqm1}) is the mechanical action of the conformal particle in a Newton-Hooke spacetime \cite{Galajinsky:2010ry}.
Notice that this mapping is valid only for a specific set of values of $n$. Moreover, it fails when both sources of curvature are present. This first case associated to $n=0$ corresponds to the original work of Pioline and Waldron \cite{Pioline:2002qz}. As one can see, the duality with the conformal particle turns out to be more general and is valid for different isotropic models such as the closed universe. Moreover, one can notice that there is a second possibility which corresponds to the two other realizations of the conformal symmetry for the (A)dS case when $n=3$ and for closed universe when $n=1$. In that case, the cosmological action can be recast into
\begin{align}
\label{accqm2}
\bar{S} [q]& =  \frac{V_0}{L^2_P} \int  \rd \eta \left[ \dot{q}^2 - \frac{\mathcal{G}}{q^2} +  \frac{\rd}{\rd \eta} \left( \Omega^2 \eta \right)  \right] 
\end{align}
where we have
\begin{align}
\Omega = \left\{
    \begin{array}{ll}
        \frac{3}{2L_{\Lambda}} & \mbox{if } L_c \rightarrow + \infty \;, \;\;n= 3 \\
        \frac{1}{L_{c}} & \mbox{if } L_{\Lambda} \rightarrow + \infty \;, \;\;n= 1 
    \end{array}
\right.
\end{align} 
and where the last term is a total derivative which does not affect the dynamics. This result shows that the (A)dS and closed universes can be mapped either onto the standard conformal particle or on its Newton-Hooke extension depending on the choice of clock used to describe their dynamics. This reflects the two possible realizations of the M\"{o}bius symmetry  for each models identified in Section~\ref{sec2}. In particular, notice that the initial result by Pioline and Waldron  \cite{Pioline:2002qz} corresponds only to one of them. Finally, we stress that in order for this general mapping to faithfully encode the cosmological dynamics, one has to assume that the conformal particle has a vanishing energy.

Before going to the next section, let us stress that this mapping onto the conformal particle might allow one to import and adapt the techniques developed for conformal quantum mechanics to quantum cosmology. Preliminary results along that line have already been reported in \cite{BenAchour:2019ufa}, following the techniques developed in \cite{Chamon:2011xk, Jackiw:2012ur}. See also \cite{Ardon:2021vae, Khodaee:2017tbk, Andrzejewski:2015jya} for more recent works on conformal quantum mechanics and the structure of its correlation functions. Another series of earlier works on conformal quantum mechanics (and more generally on systems with long-range inverse square interaction and SO$(2,1)$ symmetry) can also be found in \cite{Camblong:2005an, Camblong:2003mz, Camblong:2003mb, Ananos:2003yh, Ananos:2002id}. The results and methods developed there for these systems might therefore be adapted to investigate the quantum theory of homogeneously curved FRW geometries. In particular, the renormalization methods and the consequences of the breaking of the conformal symmetry in the strong-coupling regime might provide a new framework to discuss the physics in the nearly singular regime as well as to discuss the singularity resolution for these cosmological models. Having reviewed the different mapping and dualities, we now turn to a detailed analysis of the conserved charges generating this M\"{o}bius symmetry.



\section{Noether charges}

\label{sec3}

In this section, we focus on the $\SL(2,\mathbb{R})$ sector of the Noether symmetry and we compute the associated Noether charges. Besides being the generators of the Noether symmetry, these charges play several key roles: on one hand, they allow one to solve for the dynamics in an algebraic manner, and on the other hand, their algebra provide a key structure to guide us when quantizing these cosmological systems. We shall come back on this point in the discussion. Consider therefore the action
\begin{align}
S_n [a, \varphi]& =  \frac{V_0}{L^2_P} \int  \rd \eta \left[ \frac{L^2_P}{2} a^{3+n} \dot{\varphi}^2  -  a^{1+n} \dot{a}^2  +  a^{1-n} \left(  \frac{1}{ L^2_c}   -  \frac{ a^{2}}{L^2_{\Lambda}} \right)\right] 
\end{align}
and let us perform a infinitesimal transformation (\ref{trans1} - \ref{trans3}) where $f(\eta) = \eta + \epsilon (\eta)$. Under this infinitesimal diffeomorphism, the fields vary as
\begin{align}
\label{inf1}
\delta a & = \tilde{a} (\eta) - a (\eta) = - \epsilon \dot{a} + \lambda \dot{\epsilon} a  \\
\label{inf2}
\delta \varphi & = \tilde{a} (\varphi) - a (\varphi) = - \epsilon \dot{\varphi}
\end{align}
Upon imposing the condition $\lambda (3+n) - 1 =0$ derived above, we obtain
\begin{align}
\label{var1}
\delta S_n & =  \frac{V_0}{L^2_P} \int \rd \eta \left\{-  \frac{\rd}{\rd \eta} \left[ \epsilon \left(  \frac{L^2_P}{2} a^{3+n} \dot{\varphi}^2 - a^{1+n} \dot{a}^2 +\frac{a^{1-n}}{L^2_c} - \frac{a^{3-n}}{L^2_{\Lambda}} \right) + \frac{2\lambda}{3+n} a^{3+n}  \ddot{ \epsilon} \right]  \right. \\
\label{var2}
& \left. \qquad \qquad \qquad + \; 2 \dot{\epsilon}\;  \left[ \frac{(1-2\lambda)}{L^2_c} a^{1-n} - \frac{(1-3\lambda)}{L^2_{\Lambda}} a^{3-n}\right]  +  \; \frac{2\lambda}{3+n}a^{3+n}\dddot{\epsilon}  \right\}
\end{align}
Depending on the parameters $(L_c, L_{\Lambda})$ and on the specific values of $n$ (and thus of $\lambda$) identified in the previous section, the infinitesimal variation of the action reduces then to a total derivative
\be
\delta S_n =   \int \rd \eta \frac{\rd F_n}{\rd \eta} 
\ee
Identifying the quantity $F_n$ allows one to compute the conserved charges as follows. Using the pre-symplectic potential of the system given by
\begin{align}
\Theta_n & = \frac{2V_0}{L^2_P} \left[ \frac{L^2_P}{2} a^{3+n} \dot{\varphi} \delta \varphi -  a^{1+n} \dot{a} \delta a \right] \\
& = \frac{2V_0}{L^2_P} \left[-\lambda a^{2+n} \dot{a} \dot{\epsilon} - \epsilon \left( \frac{L^2_P}{2} a^{3+n} \dot{\varphi}^2 - a^{1+n} \dot{a}^2 \right) \right]
\end{align}
the Noether charge generating the infinitesimal transformations (\ref{inf1} - \ref{inf2}) is given by
\be
Q^{(n)} = F_n - \Theta_n
\ee
We can now evaluate them in the different cases identified above. 

\subsection{Charges for the flat model}

\label{sec3.1}

For the flat FRW case where $L_c \rightarrow + \infty$ and $L_{\Lambda} \rightarrow + \infty$, we have found that we do not need to impose any restriction on the parameter $n$ encoding the choice of proper time. The conserved charge is given by
\begin{align}
Q^{(n)} & =  \frac{V_0}{L^2_P}  \left[   \epsilon  \left( \frac{L^2_P}{2} a^{3+n} \dot{\varphi}^2 - a^{1+n} \dot{a}^2 \right)  +  \frac{2}{3+n}  a^{2+n} \dot{a} \dot{\epsilon} -  \frac{2}{(3+n)^2} a^{3+n} \ddot{\epsilon}\right] 
\end{align}
At this stage, it is instructive to consider the phase space functions
\begin{align}
H_{n} & = \frac{V_0}{L^2_P} \left( \frac{L^2_P}{2} a^{3+n} \dot{\varphi}^2 - a^{1+n} \dot{a}^2 \right)  \;, \\
C_{n} & = \frac{2V_0}{L^2_P} \frac{a^{2+n} \dot{a}}{3+n} \;, \\
\label{volconf}
v_n & = a^{3+n} \;, 
\end{align}
which correspond to the hamiltonian (which we assume to be vanishing), the dilatation operator introduced in (\ref{dil}) and finally the 3d dimensionless volume.
They satisfy the so called CVH algebra
\be
\label{CVHflat}
\{ v_n, H_n\} = \frac{(3+n)^2 L^2_P}{2V_0} \cC_n \;, \qquad \{ C_n, H_{n}\} = - H_{n} \;, \qquad \{ C_n, v_n\} = v_n
\ee
which was first discussed in \cite{BenAchour:2017qpb, BenAchour:2018jwq, BenAchour:2019ywl} in the specific case $n=0$. This structure is the key ingredient allowing to recognize directly the existence of the M\"{o}bius symmetry. Since its first appearance in \cite{BenAchour:2017qpb, BenAchour:2018jwq, BenAchour:2019ywl}, and its analysis in \cite{BenAchour:2019ufa}, it has been noticed that it actually appears in a large class of symmetry reduced models of GR, such as in Bianchi or in black hole interior models such as the Kantowski-Sachs midi-superspace \cite{Geiller:2020xze}.
In term of the CVH generators, the conserved charge reads
\begin{align}
Q^{(n)} (v, \pi_v, \pi_{\varphi} , \eta)& =    \epsilon H_{n}  +  \dot{\epsilon} \; \cC_{n}-  \frac{2V_0}{(3+n)^2 L^2_P}  \ddot{\epsilon}  v
\end{align}
Working with the standard M\"{o}bius reparametrization (\ref{mobius}), we obtain for the function $\epsilon(\eta)$ the three possibilities
\begin{align}
\epsilon = \left\{
    \begin{array}{ll}
        \sigma  & \mbox{for } \; a=d=1, \; c= 0, \; b = \sigma \;, \;  \text{translation} \\
        \sigma \eta  & \mbox{for} \; b=c=0, \; d=1, \; a = \sigma \;, \;  \text{dilatation}  \\
        \sigma \eta^2  & \mbox{for} \; b=a=0, \; d=1, \; c = \sigma \;,  \; \text{special conformal transformation}  \\
    \end{array}
\right.
\end{align}
such that $\epsilon(\eta)$ is at most quadratic in the proper time $\eta$. We have therefore $\dddot{\epsilon} =0$.
Omitting the parameter $\sigma$ which comes as global factor, the three charges generating these transformations are given by
\begin{align}
Q^{(n)}_t & = H_n \\
Q^{(n)}_d & =   \eta  H_n + C_n  \\
Q^{(n)}_s  & =  \eta^2  H_n +2 \eta C_n -  \frac{4V_0}{(3+n)^2 L^2_P} v_n 
\end{align}
Using the CVH brackets, it is then direct to check that these charges are indeed conserved in time
\be
\frac{\rd Q^{(n)}}{\rd \eta} = \{ Q^{(n)}, H_n\} + \frac{\partial Q^{(n)}}{\partial \eta} = \frac{2v }{(3+n)^2} \; \dddot{\epsilon} =0
\ee
and that they satisfy at any time an $\sl(2,\mathbb{R})$ charge algebra
\begin{align}
\{ Q^{(n)}_d, Q^{(n)}_s \} = Q^{(n)}_s \;, \qquad \{ Q^{(n)}_d, Q^{(n)}_t\} = - Q^{(n)}_t \;, \qquad \{ Q^{(n)}_t, Q^{(n)}_s \} =  2 Q^{(n)}_d 
\end{align}
These charges provide an example of evolving constants of motion for flat cosmology and their algebra reflects the fact that the proper time is defined as an $\sl(2,\mathbb{R})$ equivalence class of clocks. This result generalizes the findings of \cite{BenAchour:2019ufa} for any choice of the parameter $n \in \mathbb{Z}/\{-3\}$. Noticed that the CVH phase space functions forming the CVH algebra (\ref{CVHflat}) play the role of the initial conditions for the conserved charges $( Q^{(n)}_t, Q^{(n)}_d, Q^{(n)}_s)$. Having computed the conserved charges and their algebra for the flat model, we can express the dynamics of the volume configuration variable in term of them. Remembering that the hamiltonian has to vanish, i.e. $Q^{(n)}_t =0$, we obtain
\be
v_n (\eta) = \frac{(3+n)^2 L^2_P}{2V_0} C_n \; \eta - Q^{(n)}_s
\ee
which shows that for any choice of clock, the configuration variable evolves always linearly with the chosen clock.
We can now turn to the closed universe.

\subsection{Charges for the closed universe}

\label{sec3.2}

Consider now the model describing the closed  universe. 

\subsubsection{First clock}

In the following we shall treat the first choice of proper time corresponding to $(n, \lambda) = (+1, 1/4)$. The associated charge is given by
\begin{align}
\label{char}
Q^{+}_c = \frac{V_0}{L^2_P} \left[ \epsilon \left( \frac{L^2_P}{2} a^4 \dot{\varphi}^2 - a^2 \dot{a}^2 \right) + \frac{\dot{\epsilon}}{2}  a^3 \dot{a} - \frac{a^4}{8} \ddot{\epsilon} \right]
\end{align}
For this model and this specific choice of proper time, the CVH phase space functions read
\begin{align}
H_{+} &= \frac{V_0}{L^2_P} \left( \frac{L^2_P}{2} a^4 \dot{\varphi}^2 - a^2 \dot{a}^2 - \frac{1}{L^2_c} \right) \\
C_{+} & = \frac{V_0}{2L^2_P} a^3 \dot{a} \\
v_{+} & = a^4
\end{align}
and the CVH algebra is realized as
\be
\label{CVH2}
\{ v_{+}, H_{+}\} = \frac{ 8 L^2_P}{V_0} \cC_{+} \;, \qquad \{ C_{+}, H_{+}\} = - H_{+} - \frac{V_0}{L^2_c L^2_P}  \;, \qquad \{ C_{+}, v_{+}\} = v_{+}
\ee
Notice that in this case, the homogeneous spatial curvature is appearing as a central extension of the flat CVH algebra. Using the CVH generators, (\ref{char}) can be expressed as
\be
Q^{+}_c = \epsilon \left( H + \frac{V_0}{L^2_PL^2_c}\right) + \dot{\epsilon} \; \cC  - \frac{V_0}{8L^2_P}  \ddot{\epsilon}\; v
\ee
For this choice of proper time, the symmetry of the action is realized as standard M\"{o}bius transformation (\ref{mobius}) and one has again
\begin{align}
\epsilon (\eta)= \left\{
    \begin{array}{ll}
        \sigma  & \mbox{for } \; a=d=1, \; c= 0, \; b = \sigma \;, \;  \text{translation} \\
        \sigma \eta  & \mbox{for} \; b=c=0, \; d=1, \; a = \sigma \;, \;  \text{dilatation}  \\
        \sigma \eta^2  & \mbox{for} \; b=a=0, \; d=1, \; c = \sigma \;,  \; \text{special conformal transformation}  \\
    \end{array}
\right.
\end{align}
for the three infinitesimal operations building up the M\"{o}bius transformations. Then, the three charges generating these transformations are given by
\begin{align}
\label{t+}
Q^{+}_t & =  H_{+} + \frac{V_0}{L^2_PL^2_c}  \\
\label{d+}
Q^{+}_d & =  \eta  \left( H_{+} + \frac{V_0}{L^2_PL^2_c}\right)  + C_{+}  \\
\label{s+}
Q^{+}_s  & = \eta^2  \left( H_{+} + \frac{V_0}{L^2_PL^2_c}\right) + 2\eta C_{+} -  \frac{V_0}{4L^2_P} v_{+}
\end{align}
The conservation of the charge is straightforward and reads
\be
\frac{\rd Q^{+}_c}{\rd \eta } = - \frac{V_0}{L^2_P L^2_c} \dddot{\epsilon} =0
\ee
Moreover, one can show that these three charges satisfy also an $\sl(2, \mathbb{R})$ Lie algebra at any time given by
\begin{align}
\{ Q^{+}_d, Q^{+}_s \} = Q^{+}_s \;, \qquad \{ Q^{+}_d, Q^{+}_t\} = - Q^{+}_t \;, \qquad \{ Q^{+}_t, Q^{+}_s \} =  2 Q^{+}_d 
\end{align}
With this first choice of proper time, the constant curvature does deform the CVH structure but the charge algebra remains unchanged. Moreover, the dynamics is given by
\begin{align}
\label{tra}
v_{+}(\eta) = \frac{4}{L^2_c} \eta^2 + \frac{8L^2_P}{V_0} C_{+} \; \eta - \frac{4L^2_P Q^{+}_s}{V_0}
\end{align}
Notice that presence of the curvature through the length scale $L_c$ deforms the trajectory from a linear evolution in $\eta$ into a quadratic one.

\subsubsection{Second clock and Niederer transformations}
Consider now the second choice of proper time corresponding to $(n, \lambda) = (-1, 1/2)$. We have shown in Section~\ref{Niederer} that upon choosing this clock, the reduced action of the flat model can be mapped to the one of the closed universe using Niederer's transformations. Therefore, the dynamics as well as the conserved charges of the two models can also be mapped via such transformation. In the following, we shall therefore derive the Noether charges generating the second realization of the M\"{o}bius invariance using the Niederer's transformations. Consider thus the charges generating the dilatation and special conformal transformation for the flat model:
\begin{align}
\tilde{Q}^{-}_d  & = \frac{V_0}{L^2_P} \left[ \tilde{\eta} \left( \frac{L^2_P}{2} \tilde{a}^2  \left( \frac{ \rd \tilde{\varphi}}{\rd \tilde{\eta}} \right)^2 - \left( \frac{ \rd \tilde{a}}{\rd \tilde{\eta}} \right)^2 \right) + \tilde{a}  \frac{ \rd \tilde{a}}{\rd \tilde{\eta}} \right] \\
\tilde{Q}^{-}_s & =  \frac{V_0}{L^2_P}  \left[  \tilde{ \eta}^2  \left( \frac{L^2_P}{2} \tilde{a}^2  \left( \frac{ \rd \tilde{\varphi}}{\rd \tilde{\eta}} \right)^2 - \left( \frac{ \rd \tilde{a}}{\rd \tilde{\eta}} \right)^2 \right)  +  2   \tilde{a}  \frac{ \rd \tilde{a}}{\rd \tilde{\eta}}  \tilde{\eta} -  \tilde{a}^{2} \right] 
\end{align}
and let us apply the transformations (\ref{NT0}), (\ref{NT1}) and (\ref{NT2}). For the clock $(n, \lambda) = (-1, 1/2)$ and with $\Omega = \Omega_c = \sqrt{2}/L_c$, they read
\begin{align}
\label{NT}
\eta \rightarrow  \tilde{\eta} & = L_c \tan{\left[ \frac{\eta}{L_c}\right]} \qquad  a \rightarrow \tilde{a} = a / \cos{\left[ \frac{ \eta}{L_c}\right]}
\end{align}
while $\varphi$ transforms as a scalar. Under this transformation, the conserved charges become
\begin{align}
\label{NC1}
Q^{-}_d & = \frac{V_0}{L^2_P} \left\{  \frac{L_c}{2} \sin{\left[ \frac{2 \eta}{L_c}\right]}\left( \frac{L^2_P}{2} a^2 \dot{\varphi}^2 - \dot{a}^2 \right) + \cos{\left[ \frac{2\eta}{L_c}\right]}a\dot{a} + \sin{\left[ \frac{2 \eta}{L_c}\right]} \frac{a^2}{2L^2_c}\right\} \\
\label{NC2}
Q^{-}_s & = \frac{V_0}{L^2_P} \left\{  L^2_c \sin^2{\left[ \frac{\eta}{L_c}\right]}\left( \frac{L^2_P}{2} a^2 \dot{\varphi}^2 - \dot{a}^2 \right) + L_c \sin{\left[ \frac{2\eta}{L_c}\right]}a\dot{a} -  \cos^2{\left[ \frac{ \eta}{L_c}\right]} a^2\right\} 
\end{align} 
We can now express them in term of the CVH phase space functions of the closed universe. For this specific choice of clock $(n, \lambda) = (-1, 1/2)$, they read
\begin{align}
H_{-} &= \frac{V_0}{L^2_P} \left( \frac{L^2_P}{2} a^2 \dot{\varphi}^2 -  \dot{a}^2 - \frac{V_0}{L^2_PL^2_c} a^2 \right) \\
C_{-} & = \frac{V_0}{L^2_P} a \dot{a} \\
v_{-} & = a^2
\end{align}
and the CVH algebra is realized as
\be
\{ v_{-}, H_{-}\} = \frac{ 2 L^2_P}{V_0} \cC_{-} \;, \qquad \{ C_{-}, H_{-}\} = - H_{-} - \frac{2V_0}{L^2_c L^2_P} v \;, \qquad \{ C_{-}, v_{-}\} = v_{-}
\ee
Notice that the homogeneous spatial curvature shows up as a deformation of the bracket $ \{ C_{-}, H_{-}\}$. From the point of view of the dynamics, this deformation implies a second source of acceleration in the hamiltonian, as expected when turning on the homogeneous spatial curvature.
In term of the CVH phase space function, the charges (\ref{NC1}) and (\ref{NC2}) take the simpler form
\begin{align}
Q^{-}_t & = H_{-} \\
Q^{-}_d & =   \frac{L_c}{2} \sin{\left[ \frac{2 \eta}{L_c}\right]} H_{-}  + \cos{\left[ \frac{2\eta}{L_c}\right]} C_{-} + \sin{\left[ \frac{2 \eta}{L_c}\right]} \frac{V_0}{L^2_PL_c} v_{-} \\
Q^{-}_s & =  L^2_c \sin^2{\left[ \frac{\eta}{L_c}\right]} H_{-}  + L_c \sin{\left[ \frac{2\eta}{L_c}\right]} C_{-} -  \cos{\left[ \frac{ 2\eta}{L_c}\right]}  \frac{V_0}{L^2_P}v_{-}
\end{align}
where $[Q^{-}_t] = L^{-1}$, $[Q^{-}_d]= L$ and $[Q^{-}_s] = L^2$.
It is then straightforward to show that these charges are conserved, i.e. 
\be
\frac{\rd Q^{-}_d}{\rd \eta} = \{ Q^{-}_d, H_{-}\} + \frac{\partial Q^{-}_d}{\partial \eta} =  0  \;, \qquad \frac{\rd Q^{-}_s}{\rd \eta} =\{ Q^{-}_s, H_{-}\} + \frac{\partial Q^{-}_s}{\partial \eta} =  0
\ee
One can again solve for the evolution of the configuration variable $v_{-}$ in term of this clock to obtain
\be
v_{-} (\eta) = \frac{L^2_P}{V_0} \left[ L_c Q^{-}_d \sin{\left[ \frac{2 \eta}{L_c}\right]} -  Q^{-}_s   \cos{\left[ \frac{ 2\eta}{L_c}\right]}  \right]
\ee
which provides a periodic motion.

Let us make few remarks regarding the closed universe. A priori,  the effect of a spatially homogeneous curvature through the length scale $L_c$ in the line element (\ref{metric}) is radically different from the effects of  a cosmological constant in the dynamics. However, we have seen that both models share the very same algebraic structure and thus enjoy the same hidden symmetry when using the clocks identified in the Section~\ref{sec1}. In particular, both source of homogeneous curvature can be generated using a constant cocycle, i.e a proper time reparametrization function with a constant Schwarzian derivative. Then, the closed universe ($L_c >0$) shared the same structure as the AdS cosmology ($\Lambda <0$) while the universe with hyperbolic negatively curved spatial section ($L_c <0)$ shares the same algebraic structure as the de Sitter universe ($\Lambda >0$). To our knowledge, such relation between these cosmological models has not appeared elsewehere. The reason for such dualities comes from the existence of this hidden M\"{o}bius invariance which emerges only for some specific choices of clocks.

\subsection{Charges for the (A)dS cosmology}

\label{sec3.3}

Consider now the (A)dS cosmology. The case  $(n, \lambda) = (0, 1/3)$ has already been treated in detail in \cite{BenAchour:2020xif} and we shall therefore only provide the details on the new realization of this symmetry found in this work, namely the case $(n, \lambda) = (3, 1/6)$.  The conserved charge is given by
\begin{align}
Q = \frac{V_0}{L^2_P} \left\{ \epsilon \left( \frac{L^2_P}{2} a^6 \dot{\varphi}^2 - a^4 \dot{a}^2 \right) + \frac{\dot{\epsilon}}{3} a^5 \dot{a} - \frac{1}{18} a^6 \ddot{\epsilon} \right\}
\end{align}
For this model and this choice of proper time, the CVH observables read
\begin{align}
H & = \frac{V_0}{L^2_P} \left[ \frac{L^2_P}{2} a^6 \dot{\varphi}^2 - a^4 \dot{a}^2 + \frac{1}{L^2_{\Lambda}}\right] \\
C& = \frac{V_0}{3L^2_P} a^5 \dot{a} \\
v& = a^6
\end{align}
such that the conserved charge can be recast as
\begin{align}
Q =  \epsilon \left( H - \frac{V_0}{L^2_p L^2_{\Lambda}} \right) + \dot{\epsilon} \; C - \ddot{\epsilon} \;  \frac{V_0}{18L^2_P} v 
\end{align}
As we have seen in Section~\ref{sec2.2}, the symmetry is realized as standard M\"{o}bius reparametrization of the proper time, such that one can again extract the three charges
\begin{align}
\label{t+}
Q_t & =  H - \frac{V_0}{L^2_PL^2_{\Lambda}}  \\
\label{d+}
Q_d & =  \eta  \left( H - \frac{V_0}{L^2_PL^2_{\Lambda}}\right)  + C  \\
\label{s+}
Q_s  & = \eta^2  \left( H - \frac{V_0}{L^2_PL^2_{\Lambda}}\right) + 2\eta C -  \frac{V_0}{9L^2_P} v
\end{align}
generating respectively translation, dilatation and special conformal transformations of the proper time.
The conservation of the charges is straightforward as
\be
\frac{\rd Q}{\rd \eta } = - \frac{V_0}{L^2_P L^2_{\Lambda}} \dddot{\epsilon} =0
\ee
and one can show that these three charges satisfy also an $\sl(2, \mathbb{R})$ Lie algebra at any time given by
\begin{align}
\{ Q_d, Q_s \} = Q_s \;, \qquad \{ Q_d, Q_t\} = - Q_t \;, \qquad \{ Q_t, Q_s \} =  2 Q_d 
\end{align}
This provides therefore an alternative realization of the M\"{o}bius invariance of (A)dS cosmology than the one discussed first in \cite{Pioline:2002qz} and later on in \cite{BenAchour:2020xif}. Moreover, it parallels the discussion of Section~\ref{sec2.5} on the exietsnce of two different mappings onto the conformal particle. Finally, the dynamics of (A)dS cosmology can be written in term of this clock and the associated conserved charges as
\be
v (\eta) = \frac{18 L^2_p}{V_0} C \eta - \frac{9}{L^2_{\Lambda}} \eta^2 - \frac{9L^2_P}{V_0} Q_s
\ee 
where we have set $H=0$. Notice that for this clock, $C$ is a constant of motion. This parallels the structure found for the closed universe for which the trajectory is given by Eq~(\ref{tra}).
Having discussed in detail the Noether charges generating the M\"{o}bius invariance w.r.t the different choices of clock, we now turn to the last result of this work, namely the relation of our charges with the ones introduced by Kodama in \cite{Kodama:1979vn}.


\section{Relation to volume preserving diffeomorphisms}

\label{sec4}

As discussed in the introduction, the M\"{o}bius symmetry discussed here has been investigated in different gravitational systems and shown to belong to a larger structure still to be explored. Up to now, all the investigated models are restricted to homogeneous gravitational fields, describing either cosmological or black hole interior spacetimes. In order to go beyond, it would be desirable to see if the M\"{o}bius symmetry have some connection with more general objects which can be covariantly defined in inhomogeneous spacetimes. The goal of this section is to provide a first hint in this direction. 

In that regard, our set-up given by the line element (\ref{line}) is especially useful. Indeed, our cosmological model is described by an inhomogeneous and time-dependent metric, even if the radial dependency can be fully integrated out at the level of the action. For such spherically symmetric geometry, it is well known since the early eighties and the seminal work by Kodama that this geometry is naturally equipped with a divergence-free vector field known as the Kodama vector \cite{Kodama:1979vn}. Out of this vector field, one can build up two locally conserved currents from which one can extract the so called Kodama generators. In the following, we show that, at least for a specific choice of clock and only for the flat FRW and (A)dS models : i) one can define a third charge out of the two known Kodama charges and ii)  that these three Kodama charges are nothing else than the phase space functions forming the CVH algebra (given by Eq~(\ref{CVHflat}) and $n=0$ in the flat case). Below, we first review the Kodama's machinery and then present this interesting coincidence. 


\subsection{Review of the Kodama machinery}

\label{sec4.1}

Consider a spherically symmetric line element given by
\be
\rd s^2 = \gamma_{\mu\nu} \rd x^{\mu} \rd x^{\nu} + R^2(\tau,r) \rd\Omega^2
\ee 
where $\gamma_{\mu\nu}$ is the metric of the 2d base space with coordinates $(\tau,r)$ and $R(\tau, r)$ is the physical radius of the spacetime.
By definition, the Kodama vector is given by
\be
k^{\mu} \partial_{\mu} : = \epsilon_{\perp}^{\mu\nu} \nabla_{\nu} R(t,r)  \partial_{\mu}
\ee
where $\epsilon^{\perp}_{\mu\nu}$ is the volume form of the 2d base space with coordinates $(\tau, r)$ \cite{Kodama:1979vn}.
A key property of the Kodama vector is to be self-conserved
\be
\nabla_{\mu}k^{\mu} =0
\ee
Therefore, this vector field generates volume preserving diffeomorphisms of the geometry. Because of this, it defines automatically two conserved currents given by
\be
J^{\mu}_{+} = k^{\mu} \;, \qquad J^{\mu}_{-} = G^{\mu\nu}  k_{\nu}
\ee
Indeed, it is direct to check that $\nabla_{\mu} J^{\mu}_{+} = \nabla_{\mu} J^{\mu}_{-} =0$. Notice that these conservations laws are purely kinematical and do not require a choice of dynamics, such as the Einstein equations. They are known as the Kodama currents and they were introduced in the early 80's \cite{Kodama:1979vn}. Consider now the spacelike hypersurface $\Sigma$ with unit normal $n^{\mu} n_{\mu} = -1$ and induced metric $h_{\mu\nu} = g_{\mu\nu} + n_{\mu} n_{\nu}$. Integrating the conserved currents on this spacelike slice, we obtain the Kodama generators
\begin{align}
\label{v}
K_{+} & =  \frac{1}{L^2_P} \int_{\Sigma} \rd^3x \sqrt{h}\; n_{\mu} k^{\mu}  \\
K_{-} & = \int_{\Sigma}  \rd^3x \sqrt{h} \; n_{\mu} k_{\nu}  G^{\mu\nu} 
\end{align}
These generators are canonical quantities associated to any spherically symmetric vacuum backgrounds. Having presented briefly the Kodama machinery, it is worth pointing that since its introduction, the Kodama vector has been used for different purposes: it provides in particular an efficient tool to capture the causal structure of a given spherically symmetric spacetime and its (possibly dynamical) apparent horizons \cite{Hayward:1997jp}, formulate the thermodynamics of time-dependent spacetimes \cite{Cai:2008gw, Chen:2010sna, Cao:2010xx}, and it can also serve to select a preferred foliation where interesting simplifications occur \cite{Racz:2005pm, Csizmadia:2009dm, Abreu:2010ru}. Finally, we stress that while we should focus on the two standard generators $(K_{+}, K_{-})$ in the following, more general conserved currents can be constructed out of the Kodama vector, as discussed in detail in \cite{Abreu:2010ru}.  

Let us now come back to the two generators $(K_{+}, K_{-})$. In order to make contact with the cosmological system we are interested in, and therefore take into account a geometry filled with matter fields,
one has to extend the generator $K_{-}$ to take into account the matter sector. We consider therefore the full current $J^{\mu}_{-} = \left( L^2_P T^{\mu\nu} - G^{\mu\nu}\right) k_{\nu} $. Notice that now, the local conservation of this generator requires the imposition of the Einstein equations, such that its conservation is no longer kinematical. Integrating this current leads to the extended generator which we call again $K_{-}$ given by
\be
\label{h}
K_{-}  =  \int_{\Sigma} \rd^3x \sqrt{h} \; n_{\mu} k_{\nu} \left(   L^2_P T^{\mu\nu} - G^{\mu\nu} \right)
\ee
Notice that one could also add the contribution of the cosmological constant though a term $\Lambda g^{\mu\nu} k_{\nu}$ to the current, without spoiling its conservation. In the following, we shall restrict ourselves to $\Lambda=0$, while keeping in mind that the conclusion can be easily extended to the (A)dS case. Now, our goal is to compute the two generators $K_{\pm}$ for the flat FRW cosmology.

\subsection{From the Kodama to the CVH generators}
\label{sec4.2}

Consider  therefore the line element (\ref{met}) and let us take the limit  $L_c \rightarrow + \infty$ to remove the spatial constant curvature, giving
\be
\label{line}
\rd s^2 = - \frac{\cN^2(\tau)}{a^{2n}(\tau)}\rd \tau^2 + a^2(\tau) \left[ \rd r^2 + r^2 \rd \Omega^2\right]
\ee
At this level, we do not make a specific choice of the parameter $n$ to remain as general as possible.
We denote the metric of the three-dimensional spacelike hypersurface $\Sigma$ as
\be
\label{line}
\rd s^2_{\Sigma} =  h_{ij}\rd x^i \rd x^j =a^2(\tau) \left[ \rd r^2 + r^2 \rd \Omega^2\right]
\ee
such that its timelike unit normal $n^{\mu} n_{\mu} = -1$ is given by
\be
n_{\mu} \rd x^{\mu} =  \frac{\cN}{a^{n}} \rd \tau \;, \qquad n^{\mu} \partial_{\mu} = - \frac{a^n}{\cN} \partial_{\tau}
\ee 
Then, the physical radius being $R(\tau, r) = a(\tau)r$, the Kodama vector reads
\begin{align}
k^{\mu} \partial_{\mu} & =   \frac{a^n}{\cN }  \partial_{\tau} - \frac{\dot{a}}{\cN} a^{n-1} r \partial_{r}   \;, \qquad k_{\mu} \rd x^{\mu} =-  \frac{\cN}{a^n} \rd \tau - \frac{ \dot{a}}{\cN} a^{n+1} r \rd r 
\end{align}
for which the divergence-free property, i.e $\nabla_{\mu} k^{\mu} =0$, is easily checked.
Using these expressions, it is straightforward to notice that $n_{\mu} k^{\mu} = +1$ for any values of $n \in \mathbb{Z}$. We now can use this result to compute the first Kodama charge.

Computing the first Kodama charge, and working w.r.t the proper time $\rd \eta = \cN \rd \tau$, one obtains
\begin{align}
\label{v}
K_{+} & = \frac{1}{L^2_P} \int_{\Sigma} \rd^3x \sqrt{h}\; n_{\mu} k^{\mu}  = \frac{\V_0}{L^2_P} a^3
\end{align}
where $\V_0$ is the fiducial volume introduce in Eq~(\ref{fidvol}) (with $L_c \rightarrow + \infty$). As a result, this Kodama generator coincides with the volume configuration variable, given by Eq~(\ref{volconf}), but only for the choice of clock $n=0$, i.e. $v_{(0)} = a^3$. We can therefore write
\be
K_{+} =  \frac{V_0}{3L^2_P} v_{(0)}
\ee
To compute the second Kodama charge, we restrict ourselves to the case $n=0$. Since our matter content is a massless scalar field, one has
\be
 G^{\tau\tau} n_{\tau} k_{\tau}=   \frac{3}{\cN^2}\frac{ \dot{a}^2}{a^2}  \;, \qquad T^{\tau\tau} n_{\tau} k_{\tau} = \frac{\dot{\varphi}^2}{2\cN^2} 
\ee
Working with in term of the proper time $\rd \eta = \cN \rd \tau$, one obtains that
\begin{align}
K_{-} =  \frac{3 \V_0}{ L^2_P} \left[ \frac{L^2_P}{6} a^3 \dot{\phi}^2 - a \dot{a}^2  \right]  = \frac{V_0}{ L^2_P} \left[ \frac{L^2_P}{2} a^3 \dot{\varphi}^2 - a \dot{a}^2  \right]  = H_{(0)}
\end{align}
where we have used the rescaling (\ref{resc}). Therefore, the generators associated to the two currents introduced in \cite{Kodama:1979vn} match with the volume and hamiltonian phase space functions for the specific clock $(n, \lambda)= (0, 1/3)$. From this observation, the identification with the CVH algebra is straightforward. Using the brackets (\ref{CVHflat}) at $n=0$, one can define a third Kodama generator through the Poisson bracket of the first two generators given by
\be
K_{0} := \gamma \{ K_{-}, K_{+}\} =  C_{(0)} 
\ee
where $\gamma = 2 /3$. By construction, they satisfy the following $\sl(2,\mathbb{R})$ algebra
\be
\{ K_{+}, K_{-}\} = \frac{K_0}{\gamma} \;, \qquad \{ K_0, K_{- }\} =  - K_{+}  \;, \qquad \{ K_0, K_{+ }\} =    K_{+}
\ee
and one can construct evolving constants of motion in terms of the clock $\eta$ given by
\begin{align}
\label{kod1}
\K_t & = K_{-} \\
\label{kod2}
\K_d & =   \eta  K_{-} + K_0  \\
\label{kod3}
\K_s  & =  \eta^2  K_{-} +  2  \left( \eta  K_0  -  \frac{2}{3} K_{+}  \right)
\end{align}
These Kodama charges $\left( \K_t, \K_d, \K_s\right)$ are conserved in time and generate respectively the proper time translation, dilatation and special conformal transformations for the underlying spherically symmetric cosmology. They satisfy
\be
\{ \K_s, \K_t\} =  2 \K_d\qquad \{\K_d, \K_t \} = - \K_t \;, \qquad \{ \K_d, \K_s\} = \K_s 
\ee
Therefore, the Kodama generators $\left( K_0, K_{+}, K_{-}\right)$ can be understood, just as the CVH observables, as the initial conditions of the Noether charges $\left( \K_t, \K_d, \K_s\right)$ generating the M\"{o}bius symmetry for this specific cosmological model and clock. As emphasized above, the inclusion of a cosmological constant is straightforward and this structure generalizes to the (A)dS cosmology for the specific choice of clock $n=0$. The associated charges can be derived using again the Niederer transformation used in Section~\ref{sec3.2}. To the knowledge of the author, this is the first time the Kodama generators $\left( K_{+}, K_{-}\right)$ are related to an underlying conformal symmetry of the associated spherically symmetric gravitational system and that well defined conserved Kodama charges $\left( \K_t, \K_d, \K_s\right)$ are constructed out of them. 

This coincidence suggests that the M\"{o}bius symmetry could be related in a subtle way to volume preserving diffeomorphisms which are generated by divergence-free vector fields. However, it is worth recalling that the M\"{o}bius symmetry discussed here is a physical symmetry which maps gauge-inequivalent solutions of the equations of motion and thus is different from standard gauge transformations. Understanding the general picture requires more works and the present result raises more questions than it can answer. It might nevertheless provide an interesting starting point to investigate how the M\"{o}bius symmetry could be realized in inhomogeneous spacetimes. Up to now, the different generalizations we have attempted to reproduce the whole structure uncovered here for any $n \in \mathbb{N}/\{-3\}$ have not been successful, justifying a posteriori the restrictions used in this section. It might be that more general divergence-free vector fields can be constructed which allow one to reproduce our conserved charges for any value of $n$ and in presence of a spatial curvature, but this possibility remains to be explored and no evident solutions have appeared to us. It will be the subject of future investigations.

\section{Discussion}

\label{sec5}

We conclude with a brief summary of our results and a perspective on the future research directions. Motivated by previous results demonstrating the existence of a residual global symmetry under proper time reparametrization in different cosmological systems \cite{BenAchour:2019ufa, BenAchour:2020xif} , we have presented a unified treatment of this symmetry in a large class of homogeneous and isotropic cosmological models, completing and generalizing previous findings.

Concretely, we have considered the homogeneous and isotropic class of geometries described by the line element (\ref{metric}). This set up generalizes the previous investigations by allowing for a spatial constant curvature parametrized by the length scale $L_c$. We have focused on the Einstein-Scalar-$\Lambda$ system where the scalar matter has no self-interaction such that the symmetry reduced action is given by Eq~(\ref{ac}). This set up allows us to treat at once the flat FRW model, the (A)dS cosmology as well as the closed universe (and its counterpart with negative constant spatial curvature).  In order to provide a complete picture of the residual symmetry in this family of cosmological models, and in particular how the choice of clocks affects its realization, we have introduced a one-parameter field dependent reparametriztion of the lapse, given by Eq~(\ref{para}), where the parameter $n\in \mathbb{Z}$ encodes the infinite freedom to pick up a given internal clock to describe the dynamics. Starting from this set-up, we have shown that:
\begin{itemize}
\item Under a finite Diff$(S^1)$ transformation given by Eq~(\ref{trans1}) - Eq~(\ref{trans3}), the symmetry-reduced Einstein-Scalar-$\Lambda$ action transforms as Eq~(\ref{change}). This transformation of the action provided the conformal weight of the scale factor, namely $\lambda$, is related to the parameter $n$ through the condition (\ref{cond}). Therefore, the choice of clock dictates the value of the conformal weight. Moreover, the general formula (\ref{change}) allows us to identify in a unified manner the different realizations of the residual symmetry. From this formula, we have shown that
\begin{itemize}
 \item The flat FRW model corresponding to $L_c \rightarrow +\infty$ and $L_{\Lambda} \rightarrow + \infty$ possesses a residual symmetry under M\"{o}bius reparametrization for any value of $n \in  \mathbb{Z}/\{-3\}$. Hence, for any choice of internal clock parametrized by a given value of $n$, this symmetry is realized. This generalizes the result found in \cite{BenAchour:2019ufa} for $n=0$.
 \item When turning on the cosmological constant while keeping the spatial curvature vanishing, i.e. $L_c \rightarrow +\infty$, which corresponds to the (A)dS model, we have found that the residual symmetry is realized only for two values of $n$. For the value $n=3$, the (A)dS cosmology enjoys a M\"{o}bius symmetry similar to the flat case, while for $n=0$, the system is invariant under a reparametrization of the proper time given by Eq~(\ref{defmob}). This last case corresponds to the one treated earlier in \cite{BenAchour:2020xif}. We have therefore identified a new realization of the symmetry while recovering our previous result for the (A)dS model.
 \item When turning on only the spatial curvature while working with a vanishing cosmological constant, namely $L_{\Lambda} \rightarrow +\infty$, we find again that a residual symmetry is realized only for two specific values of the parameter $n$, and thus for two specific clocks. For $n=1$, the system is invariant under standard M\"{o}bius transformation while for $n=-1$, the system is invariant under proper time reparametrization given by Eq~(\ref{comp}). The existence of this symmetry for the model with homogeneous spatial curvature, as well as its two possible realizations, is new.
 \item Finally, the symmetry cannot be realized when both sources of (three dimensional and four dimensional) curvature are present together.
\end{itemize}
Therefore, turning on a homogeneous curvature, being three dimensional or four dimensional, reduces the number of realizations of the residual symmetry to only two clocks. What role these clocks are playing ? One can notice that, in each case, the first choice of clock allows one to recast the potential contribution as a mere constant in the Lagrangian, while the second one allows us to reabsorb it in the flat conformal anomaly (given by the cocycle). Beyond this observation, it would be interesting to understand if these specific clocks (and thus if these observers) have some additional properties. These findings show that, provided one identifies the right clocks, the residual conformal symmetry found in \cite{BenAchour:2019ufa, BenAchour:2020xif} is also a property of the model with constant spatial curvature, and that it enjoys several different realizations depending on the source of curvature which we turn on. The different possibilities are compactly captured by our approach which thus provides a unified treatment of this residual conformal symmetry for the class of homogeneous and isotropic cosmological models considered here\footnote{Other symmetries of the flat model have been recently discussed at the level of the equations of motion in \cite{Dussault:2020uvj} and it would be interesting to see how they are related to the present structure.}.
\item This surprising residual symmetry can be understood through a duality between our cosmological systems and the conformal particle introduced in \cite{deAlfaro:1976vlx}. We have shown that the duality uncovered initially by Pioline and Waldron \cite{Pioline:2002qz} is only a specific case of a more general mapping: provided one selects the right clock, the different cosmological models can be recast into the action of conformal mechanics at vanishing energy. Interestingly, the two possible realizations of the M\"{o}bius symmetry identified for each homogeneously curved models translate into two different mappings onto the conformal particle given by Eq~(\ref{accqm1}) and Eq~(\ref{accqm2}). This interesting duality opens up the possibility to import and adapt techniques developed for conformal quantum mechanics in quantum cosmology, and reconsidered standard key issues of the field from a new perspective based on symmetry arguments \cite{Ardon:2021vae, Khodaee:2017tbk, Andrzejewski:2015jya, Chamon:2011xk, Jackiw:2012ur, Camblong:2005an, Camblong:2003mz, Camblong:2003mb, Ananos:2003yh, Ananos:2002id}. Moreover, this mapping could be used as a starting point to explore how cosmological models could be related to more involved 1d conformally invariant mechanical systems, such as the well known Schwarzian mechanics \cite{Filyukov:2020iet}. 
\item Besides the M\"{o}bius symmetry transformation which leaves the cosmological action invariant, the more general finite Diff$(S^1)$ transformation shifts the action by a cocycle term which can be viewed as a new interaction term in the reduced action. The coupling constant in front of this term stands as a central charge of the system, i.e. $c = V_0/L^3_P$, which governs the conformal anomaly. It coincides with the ratio between the IR cut-off $V_0$ and the UV cut-off $L_P$, thus encoding the effective size of the homogeneous region we have considered. In the particular case where the reparametrization function has a constant Schwarzian, which corresponds to Niederer transformation, the new interaction term mimics the potential contribution of an homogeneous curvature, which allows one to generate out of the flat system the (A)dS or the spatially curved cosmological model. Therefore, this transformation can be viewed as a solution-generating map for homogeneous and isotropic cosmology. This generalizes the result of \cite{BenAchour:2020xif} and shows that there is an intriguing duality between the (A)dS model and the spatially curved one: upon choosing the suitable clock, the dynamics of both systems can be recast in the very same form thanks to their underlying conformal symmetry.
\item We have presented in detailed the Noether charges generating the different realization of the M\"{o}bius symmetry for each systems. These charges are evolving constants of motion of the system and their algebra is a key structure which fully encodes the gravitational cosmological dynamics. In particular, it allows one to solve in an algebraic manner the evolution of the configuration variable $v$. Preserving this structure at the quantum level shall provide a key guide to narrow (and even remove) standard quantum ambiguities related to factor ordering in the quantum theory. See \cite{BenAchour:2019ufa, BenAchour:2019ywl} for preliminar results in this direction. In each case, the CVH algebra built up from the hamiltonian, the volume configuration variable and the dilatation generator (which coincides with the trace of the extrinsic curvature) can be viewed as the initial conditions for the charge algebra.
\item Finally, we have used the spherically symmetric nature of our set-up to explore the status of the well known Kodama generators \cite{Kodama:1979vn}. We have shown that these generators are associated with charges which generate the residual conformal symmetry of the flat (and (A)dS) cosmological system uncovered in \cite{BenAchour:2019ufa, BenAchour:2020xif} and studied further in this work. This finding is based on the observation that the two standard Kodama generators coincide with the hamiltonian and the volume configuration variable entering in the CVH algebra, in the specific case where there is no spatial curvature and only for the choice of clock $n=0$. This observation allows one to introduce a third Kodama generator which coincides with the dilatation operator $C$ and to build via them well defined conserved charges given by Eq~(\ref{kod1})-(\ref{kod3}). While this result is restricted to a specific case, it suggests that the residual symmetry discussed here might find a generalization in term of volume preserving diffeomorphisms (which are generated by divergence-free vector fields to which the Kodama vector belongs). This further suggests possible connections with the incompressible fluid mechanics which is known to possess a similar conformal symmetry as the one discussed in this work \cite{ORaifeartaigh:2000vgz, Hassaine:1999hn, Bhattacharyya:2008kq, Horvathy:2009kz}. Whether a deeper connection exists and can be formulated in the context of the fluid/gravity correspondence is an interesting direction we expect to explore in the near future. Another interesting startegy would be to understand the structure enclosed by the Kodama generators at the level of the full theory. Results in this direction will be presented elsewhere.
\end{itemize}
In summary, our set-up has allowed us to present a complete picture of this new M\"{o}bius symmetry for a large class of homogeneous and isotropic cosmological models, completing and generalizing previous investigations \cite{BenAchour:2019ufa, BenAchour:2020xif}. As pointed out in the introduction, it has been understood that this conformal symmetry is only a corner of a deeper structure which still needs to be explore further. On the one hand, it would be interesting to identify the largest group of symmetry in these cosmological models, in which the present SL$(2,\mathbb{R})$ symmetry group is only a subsector. On the other hand, it would be interesting to understand how these symmetries generalize to more general symmetry-reduced gravitational systems such as black holes. See \cite{BenAchour:2020njq, BenAchour:2020ewm, Geiller:2020xze, Ben} for recent results in these two directions. In the end, we expect this structure to play a guiding role when attempting to describe these simplified gravitational fields from a suitable hydrodynamical limit of a more fundamental theory of quantum gravity.

Finally, let us comment on some questions left opened. The M\"{o}bius invariance discussed here stands as a residual Noether symmetry only at the level of the gauge-fixed action. It would be interesting to understand if the M\"{o}bius symmetry can be also realized at the level of the action prior to the gauge-fixing and what is the role of the lapse in this transformation. If so, what is the status of the M\"{o}bius reparametrization in term of the ungauge fixed time coordinate ? What is the role played by the lapse field ? From that perspective, it was noticed in \cite{BenAchour:2019ufa} that the M\"{o}bius transformations can be written in a different manner involving different transformations of both the scale factor and the lapse (see Eq~(2.13) in \cite{BenAchour:2019ufa}). It would be interesting to clarify the status of these alternative transformations which have remained unexplored so far, as well as the status of the charges which generate them. We expect to clarify this point in the near furture. Finally, it would also be useful to understand how this symmetry can be realized in the deparametrized picture, where the massless scalar field is used as a clock.

\section*{Acknowledgments}

The work of J. Ben Achour is supported by the Alexander Von Humboldt foundation. I am grateful to Etera Livine for the nice and enjoyable discussions over the last years. I would like also to thank Shinji Mukohyama for his constant support and Ibrahim Akal and Mohammad Ali Gorji for endless scientific and non-scientific discussions during my wonderful time in YITP.

\appendix

\section{Resolution of the differential equation}

\label{app}

In this appendix, we summarize the resolution of the third order differential equation encountered in Sections~\ref{sec2.2} and \ref{sec2.3}. Consider the ODE given for example by Eq~(\ref{eqdiff}), i.e.
\be
\label{eq}
\text{Sch}[f] + \Omega^2_c \dot{f}^2 = \Omega^2_c
\ee
where 
\be
\text{Sch}[f] := \frac{\dddot{f}}{\dot{f}} - \frac{3}{2} \left(\frac{\ddot{f}}{\dot{f}} \right)^2
\ee
In order to solve this equation, we use two properties of the Schwarzian derivative. First, the Schwarzian derivative of a function $f(\eta)$ remains invariant under a M\"{o}bius transformation such that
\be
\text{Sch}[M \circ g] = \text{Sch}[g]  \;, \qquad \text{with} \qquad M(\eta) := \frac{a \eta + b }{c \eta +b} \;, \qquad ad-bc \neq 0
\ee
Second, the Schwarzian derivative of a composed function $f = f_1 \circ f_2$ reads
\be
 \text{Sch} [ g_1 \circ g_2]=  \text{Sch}[g_2] +  \left(   \text{Sch}[g_1] \circ g_2 \right)\; \dot{g}^2_2
\ee
Using these properties of the Schwarzian derivative, solving (\ref{eq}) amounts at looking for a couple of functions $(F, f)$ such that 
\begin{align}
\text{Sch}[F] = \text{Sch}[M\circ F]  & = \Omega^2_c \\
\text{Sch}[F \circ f] &  = \Omega^2_c
\end{align}
where $\Omega_c$ is a constant pulsation. These conditions give us
\be
\text{Sch}[M \circ F] = \text{Sch}[F \circ f]
\ee
such that the solution reads 
\be
f = F^{-1} \circ M \circ F \;, 
\ee
where $M$ is a M\"{o}bius transformation and $F$ is a function admitting a constant Schwarzian derivative. For $\Omega_c >0$ which corresponds to the closed universe case, one finds
\be
F(\eta) = \tan{\left[ \frac{\Omega_c \eta}{\sqrt{2}}\right]} \;,
\ee
while for the universe with hyperbolic homogeneous curved spatial section, i.e. $\Omega_c <0$, the solution is given by the hyperbolic tangante. Therefore, the reparametrization function satisfying our condition for the closed universe reads explicitly
\be
f(\eta) := \text{Arctan}\left\{ \frac{\Omega_c}{\sqrt{2}} \left[ \frac{a \tan{\left[ \frac{\Omega_c \eta}{\sqrt{2}}\right]} + b  }{c \tan{\left[ \frac{\Omega_c \eta}{\sqrt{2}}\right]} + d}\right] \right\} \qquad \text{with} \qquad ad - bc \neq 0
\ee


\begin{thebibliography}{ab}


\bibitem{Alekseev:2010mx}
G.~A.~Alekseev,
``Thirty years of studies of integrable reductions of Einstein's field equations,''
  \href{http://arXiv.org/abs/1011.3846}{{\texttt{arXiv:1011.3846}}}
  
  \bibitem{Pioline:2002qz} 
  B.~Pioline and A.~Waldron,
  ``Quantum cosmology and conformal invariance,''
  Phys.\ Rev.\ Lett.\  {\bf 90}, 031302 (2003)
    \href{https://arxiv.org/pdf/hep-th/0209044.pdf}{{\texttt{arXiv:0209044}}}

  
  
    \bibitem{deAlfaro:1976vlx} 
  V.~de Alfaro, S.~Fubini and G.~Furlan,
  ``Conformal Invariance in Quantum Mechanics,''
  Nuovo Cim.\ A {\bf 34}, 569 (1976).
\href{http://doi:10.1007/BF02785666}{{\texttt{doi:10.1007/BF02785666}}}


  




   \bibitem{BenAchour:2019ufa} 
  J.~Ben Achour and E.~R.~Livine,
  ``Cosmology as a CFT$_1$,''
  JHEP {\bf 1912}, 031 (2019)
     \href{http://arXiv.org/abs/1909.13390}{{\texttt{arXiv:1909.13390}}}
  
  
  
    \bibitem{BenAchour:2019ywl} 
  J.~Ben Achour and E.~R.~Livine,
  ``Protected $SL(2,\mathbb{R})$ Symmetry in Quantum Cosmology,''
  JCAP {\bf 1909}, 012 (2019)
   \href{http://arXiv.org/abs/1904.06149}{{\texttt{arXiv:1904.06149}}}
   
   
     \bibitem{BenAchour:2018jwq} 
  J.~Ben Achour and E.~R.~Livine,
  ``Polymer Quantum Cosmology: Lifting quantization ambiguities using a $SL(2,\mathbb{R})$ conformal symmetry,''
  Phys.\ Rev.\ D {\bf 99}, no. 12, 126013 (2019)
   \href{http://arXiv.org/abs/1806.09290}{{\texttt{arXiv:1806.09290}}}
   
   
      \bibitem{BenAchour:2017qpb} 
  J.~Ben Achour and E.~R.~Livine,
  ``Thiemann complexifier in classical and quantum FLRW cosmology,''
  Phys.\ Rev.\ D {\bf 96}, no. 6, 066025 (2017)
      \href{http://arXiv.org/abs/1705.03772}{{\texttt{arXiv:1705.03772}}}


\bibitem{BenAchour:2020xif} 
  J.~Ben Achour and E.~R.~Livine,
  ``The cosmological constant from conformal transformations: M\"{o}bius invariance and Schwarzian action,''
  Class.\ Quant.\ Grav.\  {\bf 37}, no. 21, 215001 (2020)
\href{http://arXiv.org/abs/2004.05841}{{\texttt{arXiv:2004.05841}}}

\bibitem{Gibbons:2014zla} 
  G.~W.~Gibbons,
  ``Dark Energy and the Schwarzian Derivative,''
  \href{http://arXiv.org/abs/1403.5431}{{\texttt{arXiv:1403.5431}}}

\bibitem{Alon:2020yjq} 
  A.~E.~Faraggi and M.~Matone,
  ``The Geometrical Origin of Dark Energy,''
  Eur.\ Phys.\ J.\ C {\bf 80}, no. 11, 1094 (2020)
  \href{http://arXiv.org/abs/2006.11935}{{\texttt{arXiv:2006.11935}}}

\bibitem{BenAchour:2020njq} 
  J.~Ben Achour and E.~R.~Livine,
  ``Conformal structure of FLRW cosmology: spinorial representation and the $ \mathfrak{so} $ (2, 3) algebra of observables,''
  JHEP {\bf 2003}, 067 (2020)
    \href{http://arXiv.org/abs/2001.11807}{{\texttt{arXiv:2001.11807}}}
  
      \bibitem{BenAchour:2020ewm} 
  J.~Ben Achour and E.~R.~Livine,
  ``Cosmological spinor,''
  Phys.\ Rev.\ D {\bf 101}, no. 10, 103523 (2020)
  \href{http://arXiv.org/abs/2004.06387}{{\texttt{arXiv:2004.06387}}}

          \bibitem{Geiller:2020xze} 
  M.~Geiller, E.~R.~Livine and F.~Sartini,
  ``Symmetries of the Black Hole Interior and Singularity Regularization,''
  SciPost Phys.\  {\bf 10}, 022 (2021)
\href{http://arXiv.org/abs/2010.07059}{{\texttt{arXiv:2010.07059}}}.

      \bibitem{Ben} 
  J.~Ben Achour and E.~R.~Livine,
  ``Symmetries and conformal bridge in Schwarzscild-(A)dS black hole mechanics''
 (2021)
  \href{http://arXiv.org/abs/2110.01455}{{\texttt{arXiv:2110.01455}}}
  
  
 \bibitem{Nied} 
U.~Niederer, 
''The maximal kinematical invariance group of the harmonic oscillator,'' Helv. Phys. Acta 46 (1973) 191

 \bibitem{Galajinsky:2010ry}  
  A.~Galajinsky,
  ``Conformal mechanics in Newton-Hooke spacetime,''
  Nucl.\ Phys.\ B {\bf 832}, 586 (2010)
     \href{http://arXiv.org/abs/1002.2290}{{\texttt{arXiv:1002.2290}}}.
     
      \bibitem{Evol}  
     C. Rovelli, ''Time in Quantum Gravity: A hypothesis", Phys. Rev. D43, 442 (1991).
     
\bibitem{Anderson:1995tu}
A.~Anderson,
``Evolving constants of motion,''
     \href{http://arXiv.org/abs/9507038}{{\texttt{arXiv:9507038}}}.

  
\bibitem{Kodama:1979vn} 
  H.~Kodama,
  ``Conserved Energy Flux for the Spherically Symmetric System and the Back Reaction Problem in the Black Hole Evaporation,''
  Prog.\ Theor.\ Phys.\  {\bf 63}, 1217 (1980).
     \href{https://academic.oup.com/ptp/article/63/4/1217/1853503}{{\texttt{doi:10.1143/PTP.63.1217}}}.
    
     
     \bibitem{Hayward:1997jp}
S.~A.~Hayward,
``Unified first law of black hole dynamics and relativistic thermodynamics,''
Class. Quant. Grav. \textbf{15}, 3147-3162 (1998)
  \href{https://arxiv.org/pdf/gr-qc/9710089.pdf}{{\texttt{arXiv:9710089}}}.
  
       \bibitem{Cai:2008gw}
R.~G.~Cai, L.~M.~Cao and Y.~P.~Hu,
``Hawking Radiation of Apparent Horizon in a FRW Universe,''
Class. Quant. Grav. \textbf{26}, 155018 (2009)
  \href{http://arXiv.org/abs/0809.1554}{{\texttt{arXiv:0809.1554}}}.
      
       
       \bibitem{Chen:2010sna}
Y.~X.~Chen, J.~L.~Li and Y.~Q.~Wang,
``Thermodynamics for Kodama observer in general spherically symmetric spacetimes,'' (2010)
\href{http://arXiv.org/abs/1008.3215}{{\texttt{arXiv:1008.3215}}}.

\bibitem{Cao:2010xx}
Q.~J.~Cao, Y.~X.~Chen and K.~N.~Shao,
``Clausius relation and Friedmann equation in FRW universe model,''
JCAP \textbf{05}, 030 (2010)
\href{http://arXiv.org/abs/1001.2597}{{\texttt{arXiv:1001.2597}}}.

     
 \bibitem{Racz:2005pm}
I.~Racz,
``On the use of the Kodama vector field in spherically symmetric dynamical problems,''
Class. Quant. Grav. \textbf{23}, 115-124 (2006)
       \href{https://arxiv.org/pdf/gr-qc/0511052.pdf}{{\texttt{arXiv:0511052}}}.
       
       \bibitem{Csizmadia:2009dm}
P.~Csizmadia and I.~Racz,
``Gravitational collapse and topology change in spherically symmetric dynamical systems,''
Class. Quant. Grav. \textbf{27}, 015001 (2010)
  \href{http://arXiv.org/abs/0911.2373}{{\texttt{arXiv:0911.2373}}}.

\bibitem{Abreu:2010ru} 
  G.~Abreu and M.~Visser,
  ``Kodama time: Geometrically preferred foliations of spherically symmetric spacetimes,''
  Phys.\ Rev.\ D {\bf 82}, 044027 (2010)
       \href{http://arXiv.org/abs/1004.1456}{{\texttt{arXiv:1004.1456}}}.
       
       \bibitem{Ghosh:2001an}
P.~K.~Ghosh,
``Conformal symmetry and the nonlinear Schrodinger equation,''
Phys. Rev. A \textbf{65}, 012103 (2002)
 \href{https://arxiv.org/pdf/cond-mat/0102488.pdf}{{\texttt{arXiv:0102488}}}.

       
       
\bibitem{Lidsey:2018byv} 
  J.~E.~Lidsey,
  ``Inflationary Cosmology, Diffeomorphism Group of the Line and Virasoro Coadjoint Orbits,''
   \href{http://arXiv.org/abs/1802.09186}{{\texttt{arXiv:1802.09186}}}.
       
       \bibitem{Gielen:2013naa} 
  S.~Gielen, D.~Oriti and L.~Sindoni,
  ``Homogeneous cosmologies as group field theory condensates,''
  JHEP {\bf 1406}, 013 (2014)
     \href{http://arXiv.org/abs/1311.1238}{{\texttt{arXiv:1311.1238}}}.
       
       \bibitem{Oriti:2016qtz} 
  D.~Oriti, L.~Sindoni and E.~Wilson-Ewing,
  ``Emergent Friedmann dynamics with a quantum bounce from quantum gravity condensates,''
  Class.\ Quant.\ Grav.\  {\bf 33}, no. 22, 224001 (2016)
       \href{http://arXiv.org/abs/1602.05881}{{\texttt{arXiv:1602.05881}}}.
  
  \bibitem{Oriti:2016ueo} 
  D.~Oriti, L.~Sindoni and E.~Wilson-Ewing,
  ``Bouncing cosmologies from quantum gravity condensates,''
  Class.\ Quant.\ Grav.\  {\bf 34}, no. 4, 04LT01 (2017)
         \href{http://arXiv.org/abs/1602.08271}{{\texttt{arXiv:1602.08271}}}.
  
  \bibitem{Oriti:2016acw} 
  D.~Oriti,
  ``The universe as a quantum gravity condensate,''
  Comptes Rendus Physique {\bf 18}, 235 (2017)
           \href{http://arXiv.org/abs/1612.09521}{{\texttt{arXiv:1612.09521}}}.
           
           \bibitem{Pithis:2019tvp} 
  A.~G.~A.~Pithis and M.~Sakellariadou,
 ``Group field theory condensate cosmology: An appetizer,''
  Universe {\bf 5}, no. 6, 147 (2019)
    \href{http://arXiv.org/abs/1904.00598}{{\texttt{arXiv:1904.00598}}}.
       
       
       \bibitem{Rangamani:2009xk}
M.~Rangamani,
``Gravity and Hydrodynamics: Lectures on the fluid-gravity correspondence,''
Class. Quant. Grav. \textbf{26}, 224003 (2009)
           \href{http://arXiv.org/abs/0905.4352}{{\texttt{arXiv:0905.4352}}}.
           
  \bibitem{Hoehn:2011jw}
P.~A.~Hoehn,
``Effective relational dynamics,''
J. Phys. Conf. Ser. \textbf{360}, 012014 (2012)
 \href{http://arXiv.org/abs/1110.5631}{{\texttt{arXiv:1110.5631}}}.
        

\bibitem{Ardon:2021vae}
R.~d.~Ard\'on,
``Conformal quantum mechanics as a Floquet-Dirac system,''
 \href{http://arXiv.org/abs/2103.15248}{{\texttt{arXiv:2103.15248}}}.

 
 
         \bibitem{Khodaee:2017tbk}
S.~Khodaee and D.~Vassilevich,
``Note on correlation functions in conformal quantum mechanics,''
Mod. Phys. Lett. A \textbf{32}, no.29, 1750157 (2017)
        \href{http://arXiv.org/abs/1706.10225}{{\texttt{arXiv:1706.10225}}}
  
        \bibitem{Andrzejewski:2015jya}
K.~Andrzejewski,
``Quantum conformal mechanics emerging from unitary representations of SL(2,$\mathbb{R}$),''
Annals Phys. \textbf{367}, 227-250 (2016)
        \href{http://arXiv.org/abs/1506.05596}{{\texttt{arXiv:1506.05596}}}
 
             \bibitem{Chamon:2011xk} 
  C.~Chamon, R.~Jackiw, S.~Y.~Pi and L.~Santos,
  ``Conformal quantum mechanics as the CFT$_1$ dual to AdS$_2$,''
  Phys.\ Lett.\ B {\bf 701}, 503 (2011)
      \href{http://arXiv.org/abs/1106.0726}{{\texttt{arXiv:1106.0726}}}
  
  \bibitem{Jackiw:2012ur} 
  R.~Jackiw and S.-Y.~Pi,
  ``Conformal Blocks for the 4-Point Function in Conformal Quantum Mechanics,''
  Phys.\ Rev.\ D {\bf 86}, 045017 (2012)
  Erratum: [Phys.\ Rev.\ D {\bf 86}, 089905 (2012)]
        \href{http://arXiv.org/abs/1205.0443}{{\texttt{arXiv:1205.0443}}}
        
        
          \bibitem{Camblong:2005an} 
  H.~E.~Camblong, L.~N.~Epele, H.~Fanchiotti, C.~A.~Garcia Canal and C.~R.~Ordonez,
  ``Effective field theory program for conformal quantum anomaly,''
  Phys.\ Rev.\ A {\bf 72}, 032107 (2005)
\href{https://arxiv.org/pdf/hep-th/0501193.pdf}{{\texttt{arXiv:0501193}}}
  
  \bibitem{Camblong:2003mz} 
  H.~E.~Camblong and C.~R.~Ordonez,
  ``Anomaly in conformal quantum mechanics: From molecular physics to black holes,''
  Phys.\ Rev.\ D {\bf 68}, 125013 (2003)
  \href{https://arxiv.org/pdf/hep-th/0303166.pdf}{{\texttt{arXiv:0303166}}}
  
  \bibitem{Camblong:2003mb} 
  H.~E.~Camblong and C.~R.~Ordonez,
  ``Renormalization in conformal quantum mechanics,''
  Phys.\ Lett.\ A {\bf 345}, 22 (2005)
    \href{https://arxiv.org/pdf/hep-th/0305035.pdf}{{\texttt{arXiv:0305035}}}
  
  \bibitem{Ananos:2003yh} 
  G.~N.~J.~Ananos, H.~E.~Camblong and C.~R.~Ordonez,
  ``SO(2,1) conformal anomaly: Beyond contact interactions,''
  Phys.\ Rev.\ D {\bf 68}, 025006 (2003)
      \href{https://arxiv.org/pdf/hep-th/0302197.pdf}{{\texttt{arXiv:0302197}}}
  
  \bibitem{Ananos:2002id} 
  G.~N.~J.~Ananos, H.~E.~Camblong, C.~Gorrichategui, E.~Hernadez and C.~R.~Ordonez,
  ``Anomalous commutator algebra for conformal quantum mechanics,''
  Phys.\ Rev.\ D {\bf 67}, 045018 (2003)
        \href{https://arxiv.org/pdf/hep-th/0205191.pdf}{{\texttt{arXiv:0205191}}}
        
        







       
    
 
    

  
  
  \bibitem{Andrzejewski:2014boa} 
  K.~Andrzejewski,
  ``Conformal Newton-Hooke algebras, Niederer's transformation and Pais-Uhlenbeck oscillator,''
  Phys.\ Lett.\ B {\bf 738}, 405 (2014)
   \href{http://arXiv.org/abs/1409.3926}{{\texttt{arXiv:1409.3926}}}
  
   \bibitem{Galajinsky:2015xla} 
  A.~Galajinsky and I.~Masterov,
  ``On dynamical realizations of l-conformal Galilei and Newton-Hooke algebras,''
  Nucl.\ Phys.\ B {\bf 896}, 244 (2015)
     \href{http://arXiv.org/abs/1503.08633}{{\texttt{arXiv:1503.08633}}}
  
  \bibitem{Filyukov:2020iet} 
  S.~Filyukov and I.~Masterov,
  ``On the Schwarzian counterparts of conformal mechanics,''
       \href{http://arXiv.org/abs/2004.03304}{{\texttt{arXiv:2004.03304}}}
  
     
   
\bibitem{Lindblad:1969zz} 
  G.~Lindblad and B.~Nagel,
  ``Continuous Bases For Unitary Irreducible Representations Of Su(1,1),''
  PRINT-69-1210.
  
  \bibitem{Dussault:2020uvj} 
  S.~Dussault and V.~Faraoni,
  ``A new symmetry of the spatially flat Einstein-Friedmann equations,''
  Eur.\ Phys.\ J.\ C {\bf 80}, no. 11, 1002 (2020)
         \href{http://arXiv.org/abs/2009.03235}{{\texttt{arXiv:2009.03235}}}
         
         \bibitem{ORaifeartaigh:2000vgz} 
  L.~O'Raifeartaigh and V.~V.~Sreedhar,
  ``The Maximal kinematical invariance group of fluid dynamics and explosion - implosion duality,''
  Annals Phys.\  {\bf 293}, 215 (2001)
   \href{https://arxiv.org/pdf/hep-th/0007199.pdf}{{\texttt{arXiv:0007199}}}
   
   \bibitem{Hassaine:1999hn}
M.~Hassaine and P.~A.~Horvathy,
``Field dependent symmetries of a nonrelativistic fluid model,''  
Annals Phys. \textbf{282}, 218-246 (2000)
   \href{https://arxiv.org/pdf/math-ph/9904022}{{\texttt{arXiv:9904022}}}

  
    \bibitem{Bhattacharyya:2008kq} 
  S.~Bhattacharyya, S.~Minwalla and S.~R.~Wadia,
  ``The Incompressible Non-Relativistic Navier-Stokes Equation from Gravity,''
  JHEP {\bf 0908}, 059 (2009)
    \href{http://arXiv.org/abs/0810.1545}{{\texttt{arXiv:0810.1545}}}
  
  \bibitem{Horvathy:2009kz} 
  P.~A.~Horvathy and P.-M.~Zhang,
  ``Non-relativistic conformal symmetries in fluid mechanics,''
  Eur.\ Phys.\ J.\ C {\bf 65}, 607 (2010)
      \href{http://arXiv.org/abs/0906.3594}{{\texttt{arXiv:0906.3594}}}
  

  

  
   

  

  
  
  
  

    

\end{thebibliography}
\end{document}